\pdfoutput=1

\documentclass[11pt,twoside,a4paper,cmspaper,final,collab]{cms-tdr}
\def\svnVersion{475032P}\def\cmsCernNoTag{CERN-EP-2018-129}\def\cmsCernDate{\today}\def\cmsMessage{Published in the Journal of High Energy Physics as \href{http://dx.doi.org/10.1007/JHEP09(2018)046}{\doi{10.1007/JHEP09(2018)046.}}}

\begin{document}\cmsNoteHeader{EXO-16-055}

\hyphenation{had-ron-i-za-tion}
\hyphenation{cal-or-i-me-ter}
\hyphenation{de-vices}
\RCS$HeadURL: svn+ssh://svn.cern.ch/reps/tdr2/papers/EXO-16-055/trunk/EXO-16-055.tex $
\RCS$Id: EXO-16-055.tex 472327 2018-08-20 18:01:12Z alverson $

\newlength\cmsFigWidth
\setlength\cmsFigWidth{0.4\textwidth}
\providecommand{\cmsLeft}{left\xspace}
\providecommand{\cmsRight}{right\xspace}
\newlength\cmsTabSkip\setlength{\cmsTabSkip}{1ex}

\newcommand{\Hgg}{\ensuremath{\Ph\to\gamma\gamma}\xspace}
\newcommand{\Hbb}{\ensuremath{\Ph\to\bbbar\xspace}\xspace}
\newcommand{\Htt}{\ensuremath{\Ph\to\Pgt^+\Pgt^-}\xspace}
\newcommand{\wjet}{\ensuremath{\PW+\text{jets}}\xspace}
\newcommand{\zjet}{\ensuremath{\PZ+\text{jets}}\xspace}
\newcommand{\mttot}{\ensuremath{M_\mathrm{T}^{\text{tot}}}\xspace}
\newcommand{\mgg}{\ensuremath{m_{\gamma\gamma}}\xspace}
\newcommand{\ptgg}{\ensuremath{p_{\mathrm{T}\gamma\gamma}}\xspace}
\newcommand{\mA}{\ensuremath{m_{\PSA}}}
\newcommand{\mZ}{\ensuremath{m_{\PZpr}}}
\newcommand{\mDM}{\ensuremath{m_{\mathrm{DM}}}}
\newcommand{\mMed}{\ensuremath{m_{\mathrm{ med}}}}
\newcommand{\gDM}{\ensuremath{g_{\mathrm{DM}}}}
\newcommand{\gq}{\ensuremath{g_{\cPq}}}
\newcommand{\gZp}{\ensuremath{g_{\PZpr}}}
\newcommand{\PhB}{\ensuremath{\Ph_{\mathrm{B}}}\xspace}
\newcommand{\Ae}{\ensuremath{\mathrm{A}\epsilon}\xspace}
\newcommand{\SigSI}{\ensuremath{\sigma^{\mathrm{SI}}}\xspace}
\newcommand{\SigCL}{\ensuremath{\sigma_{\mathrm{95\% \,\CL}}}\xspace}
\newcommand{\SigTH}{\ensuremath{\sigma_{\mathrm{th}}}\xspace}
\providecommand{\NA}{\ensuremath{\text{---}}}
\providecommand{\cmsTable}[1]{\resizebox{\textwidth}{!}{#1}}

\cmsNoteHeader{EXO-16-055}

\title{Search for dark matter produced in association with a Higgs boson decaying to $\gamma\gamma$ or $\tau^+\tau^-$ at $\sqrt{s} = 13\TeV$}

\date{\today}

\abstract{
A search for dark matter particles is performed by looking for events with large transverse momentum imbalance and a recoiling Higgs boson decaying
to either a pair of photons or a pair of $\Pgt$ leptons.
The search is based on proton-proton collision data at a center-of-mass energy of 13\TeV collected at the CERN LHC in 2016
and corresponding to an integrated luminosity of $35.9\fbinv$.
No significant excess over the expected standard model background is observed.
Upper limits at 95\% confidence level are presented for the product of the production cross section and branching fraction in the context of two benchmark simplified models.
For the \PZpr-two-Higgs-doublet model (where \PZpr is a new massive boson mediator) with an intermediate heavy pseudoscalar particle of mass $\mA = 300\GeV$ and $\mDM = 100\GeV$,
the \PZpr masses from 550\GeV to 1265\GeV are excluded.
For a baryonic \PZpr model, with $\mDM = 1\GeV$, \PZpr masses up to 615\GeV are excluded.
Results are also presented for the spin-independent cross section for the dark matter-nucleon interaction
as a function of the mass of the dark matter particle.
This is the first search for dark matter particles produced in association with a Higgs boson decaying to two $\Pgt$ leptons.
}

\hypersetup{
pdfauthor={CMS Collaboration},
pdftitle={Search for dark matter produced in association with a Higgs boson decaying to gamma gamma or tau tau at sqrt(s) = 13 TeV},
pdfsubject={CMS},
pdfkeywords={CMS, physics, dark matter, Higgs boson, mono-Higgs}}

\maketitle

\section{Introduction\label{sec:Introduction}}

Astrophysical evidence strongly suggests the existence of dark matter (DM) in the universe~\cite{Planck}.
Whether the DM has a particle origin remains a mystery~\cite{FNAL_review}.
There are a number of well-motivated theories beyond the standard model (SM) of particle physics
that predict the existence of a particle, \PGc, that could serve as a DM candidate.
To date, only gravitational interactions between DM and SM particles have been observed.
However, the discovery of a Higgs boson by both the ATLAS and CMS Collaborations at the
CERN LHC in 2012~\cite{HiggsObs_ATLAS,HiggsObs_CMS,HiggsObs_CMS_Long}
provides a new way to probe DM-SM particle interactions.

Collider experiment searches have typically looked for DM recoiling against an associated SM particle.
Since any produced DM is unlikely to interact with the detector material, it creates an imbalance in the recorded momentum
yielding a large amount of missing transverse momentum, \ptmiss.
This paper presents a search for DM recoiling against an SM-like Higgs boson (\Ph) using the $\Ph+\ptmiss$ signature.
This SM-like Higgs boson can be produced from initial- or final-state radiation, or from a new interaction between DM and SM particles.
However, initial-state radiation of an SM-like Higgs boson from a quark or gluon is suppressed by Yukawa or loop processes, respectively~\cite{monoHiggs3,monoHiggs2,monoHiggs1}.

Previous searches for $\Ph+\ptmiss$ have been performed at both the ATLAS and CMS experiments.
No excesses were observed in either \Hbb or \Hgg decay channels
with 20.3\,(36)\fbinv of data at $\sqrt{s} = 8\,(13)\TeV$~\cite{ATLAS_MonoHbb_8TeV,ATLAS_MonoHgg_8TeV,Aaboud:2017yqz}
or with 2.3--36.1\fbinv of data at $\sqrt{s}=13\TeV$~\cite{ATLAS_MonoHbb_13TeV,CMS_MonoH_13TeV,ATLAS_MonoHgg_13TeV}.
This paper examines two Higgs boson decay channels: \Hgg and \Htt.

\begin{figure}[hbpt]
  \centering
  \includegraphics[width=\cmsFigWidth]{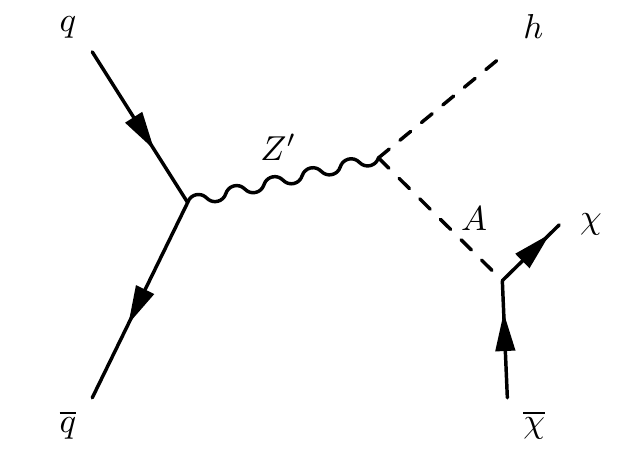}
  \includegraphics[width=\cmsFigWidth]{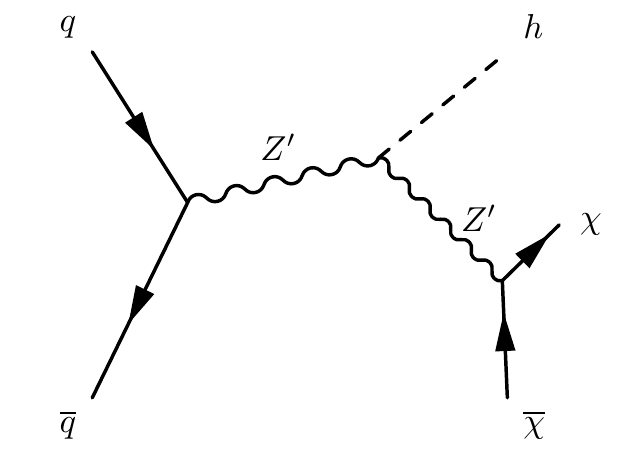}
  \caption{Leading order Feynman diagrams for DM associated production with a Higgs boson
for two theoretical models: \PZpr-2HDM (left) and baryonic \PZpr (right).}
  \label{fig:diagrams}
\end{figure}

Two simplified models for DM+\Ph production are used as benchmarks for this search,
both of which were recommended by the LHC Dark Matter Forum~\cite{darkMatterForum}.
The leading order (LO) Feynman diagrams for these models are shown in Fig.~\ref{fig:diagrams}.
The first benchmark model (Fig.~\ref{fig:diagrams} left) is a \PZpr-two-Higgs-doublet model (\PZpr-2HDM)~\cite{monoHiggs2}.
In this scenario, the SM is extended by a $U(1)_{\PZpr}$ group, with a new massive \PZpr boson mediator, while
a Type-2 2HDM framework~\cite{Craig:2013hca,Branco:2011iw} is used to formulate the extended Higgs sector.
At LO, the \PZpr boson is produced resonantly and decays into an SM-like Higgs boson and an intermediate heavy
pseudoscalar particle (\PSA). The \PSA then decays into a pair of Dirac fermionic DM particles.
This analysis does not consider the contribution of the decay $\PZpr\to\PZ\Ph$ which can have a $\Ph+\ptmiss$ signature if $\PZ\to\nu\nu$.
The second diagram (Fig.~\ref{fig:diagrams} right) describes a baryonic \PZpr model~\cite{monoHiggs1}.
In this scenario, a new massive vector mediator \PZpr emits a Higgs boson and then decays to a pair of Dirac fermionic DM particles.
Here, the baryonic gauge boson \PZpr arises from a new $U(1)_{\mathrm{B}}$ baryon number symmetry.
A baryonic Higgs boson (\PhB) is introduced to spontaneously break the new symmetry and generates the \PZpr boson mass via a coupling that is dependent on the \PhB vacuum expectation value.
The \PZpr couplings to quarks and DM are proportional to the $U(1)_{\mathrm{B}}$ gauge couplings.
There is a mixing between \PhB and the SM Higgs boson, allowing the \PZpr to radiate an SM-like Higgs boson.
The stable baryonic states in this model are the candidate DM particles.

In the \PZpr-2HDM, there are several parameters that affect the predicted cross section.
However, when the \PSA is on-shell, only the \PZpr and \PSA masses affect the kinematic distributions of
the final state particles studied in this analysis.
This paper considers a \PZpr resonance with mass between 450 and 2000\GeV and an \PSA pseudoscalar with mass between 300 and 700\GeV, in accordance with the LHC Dark Matter Forum recommendations~\cite{darkMatterForum}.
The ratio of the vacuum expectation values of the Higgs doublets ($\tan{\beta}$) in this model is fixed to 1.
As given in Ref.~\cite{CMS_MonoH_13TeV}, the DM particle mass is fixed to 100\GeV, the DM-\PSA coupling strength \gDM{}  is fixed to 1,
and the \PZpr coupling strength $\gZp$ is fixed to 0.8.

For the baryonic \PZpr model, this paper considers a \PZpr resonance with a mass between 100 and 2500\GeV and DM particle masses between 1 and 900\GeV.
As suggested for this model~\cite{presentDM}, the mediator-DM coupling is fixed to 1 and the mediator-quark coupling (\gq) is fixed to 0.25.
The mixing angle between the baryonic Higgs boson and the SM-like Higgs boson is set to 0.3 and the coupling between
the \PZpr boson and the SM-like Higgs boson is proportional to the mass of the \PZpr boson.

For both models, values of the couplings and mixing angle are chosen to maximize the predicted cross section.
Results for other values can be obtained by rescaling the cross section since these parameters do not affect the kinematic
distributions of the final state particles.
The SM-like Higgs boson is assumed to be the already observed 125\GeV Higgs boson,
since the SM-like Higgs boson has similar properties to the SM Higgs boson.
Therefore, in this paper the observed 125\GeV Higgs boson is denoted by \Ph.

Although the SM Higgs boson branching fractions to $\gamma\gamma$ and $\Pgt^+\Pgt^-$ are smaller than the branching fraction to $\bbbar$,
the analysis presented here exploits these two decay channels because they have unique advantages compared with the \Hbb channel.
The \Hgg channel benefits from higher precision in reconstructed invariant mass
and the \Htt channel benefits from smaller SM background.
Additionally, the \Hgg and \Htt channels are not dependent on \ptmiss trigger thresholds,
as such searches in these channels are complementary to those in the \Hbb channel since they can probe DM scenarios with lower \ptmiss.
The search in the \Hgg channel uses a fit in the diphoton invariant mass spectrum to extract the signal yield.
In addition to a high-\ptmiss category, a low-\ptmiss category is also considered to extend the phase space of the search.
In the \Htt channel, the three decay channels of the $\tau$ lepton with the highest branching fractions are analyzed.
After requiring an amount of \ptmiss in order to sufficiently suppress the quantum chromodynamic (QCD) multijet background,
the signal is extracted by performing a simultaneous fit to the transverse mass of the \ptmiss and the two $\tau$ lepton candidates
in the signal region (SR) and control regions (CRs).

The paper is organized as follows.
Section~\ref{sec:CMSDetector} gives a brief description of the CMS detector and the event reconstruction.
Section~\ref{sec:Samples} details the data set and the simulated samples used in the analysis.
Then Sections~\ref{sec:AnalysisHgg} and \ref{sec:AnalysisHtt} present the event selection and analysis strategy for each decay channel, respectively.
The systematic uncertainties affecting the analysis are presented in Section~\ref{sec:Systematics}.
Section~\ref{sec:Results} details the results of the analysis and their interpretations.
A summary is given in Section~\ref{sec:Conclusion}.

\section{The CMS detector and event reconstruction\label{sec:CMSDetector}}

The central feature of the CMS detector is a superconducting solenoid, of 6\unit{m} internal diameter,
providing an axial magnetic field of 3.8\unit{T} along the beam direction.
Within the solenoid volume are a silicon pixel and strip tracker, a lead-tungstate crystal electromagnetic calorimeter (ECAL),
and a brass and scintillator hadron calorimeter (HCAL).
Extensive forward calorimetry complements the coverage provided by the barrel and endcap detectors.
Charged particle trajectories are measured by the silicon pixel
and strip tracker system, covering $0 \le \phi \le 2\pi$ in azimuth and $\abs{\eta}<2.50$, where the pseudorapidity is $\eta = - \ln{\left(\tan{\theta/2}\right)}$,
and $\theta$ is the polar angle with respect to the counterclockwise beam direction.
Muons are measured in gas-ionization detectors embedded in the steel flux-return yoke.
A more detailed description of the CMS detector can be found in Ref.~\cite{Chatrchyan:2008zzk}.

Events of interest are selected using a two-tiered trigger system~\cite{Khachatryan:2016bia}.
The first level, composed of custom hardware processors, uses information from the calorimeters and muon detectors
to select events at a rate of around 100\unit{kHz} within a time interval of less than 4\mus.
The second level, known as the high-level trigger, consists of a farm of processors running a version of the full event
reconstruction software optimized for fast processing, and reduces the event rate to around 1\unit{kHz} before data storage.

Using information from all CMS subdetectors, a global event reconstruction is performed using the particle-flow (PF) algorithm~\cite{CMS-PRF-14-001}.
The PF algorithm optimally combines all of the detector information and generates a list of stable particles (PF candidates),
namely photons, electrons, muons, and charged and neutral hadrons.
The reconstructed vertex with the largest value of summed physics-object $\pt^2$ is taken to be the primary $\Pp\Pp$ interaction vertex (PV).
The physics objects are the jets, clustered using the jet finding algorithm~\cite{Cacciari:2008gp,Cacciari:fastjet1} with the tracks assigned
to the vertex as inputs, and the negative vector sum of the \pt of those jets.
The PV is used as the reference vertex for all objects reconstructed with the PF algorithm.

Photons are reconstructed from their energy deposits in the ECAL, which can involve several crystals~\cite{egamma_phoReco8TeV}.
A photon that converts to an electron-positron pair in the tracker will yield a shower spread out in azimuth due to the deflection of the electron and positron in the strong magnetic field.
In order to achieve the best photon energy resolution, corrections are applied to overcome energy losses
including those from photon conversions~\cite{egamma_phoReco8TeV}.
Additional corrections, calculated from the mass distribution of $\PZ \to \Pe^+\Pe^-$ events,
are applied to the measured energy scale of the photons in data (${\le}1$\%)
and to the energy resolution in simulation (${\le}2$\%).

Electron reconstruction requires the matching of the cluster of energy deposits in the ECAL with a track in the silicon tracker.
Electron identification is based on the ECAL shower shape, matching between the track and ECAL cluster, and consistency with the PV.
Muons are reconstructed by combining two complementary algorithms~\cite{CMSMuonJINST}:
one that matches tracks in the silicon tracker with signals in the muon system,
and another in which a global track fit seeded by the muon track segment is performed.

Jets are reconstructed from PF candidates using the anti-\kt clustering algorithm~\cite{Cacciari:2008gp}
as implemented in \FASTJET~\cite{Cacciari:fastjet1} with a distance parameter of 0.4.
Jet energy corrections are derived from simulation to bring the average measured response of jets to that of particle-level jets.
Hadronically decaying $\tau$ leptons are reconstructed from jets using the hadrons-plus-strips (HPS) algorithm~\cite{hpssource}.
The HPS algorithm uses combinations of reconstructed charged hadrons and energy deposits in the ECAL to reconstruct the $\tau$ lepton's three most common
hadronic decay modes: 1-prong, 1-prong + $\PGpz\mathrm{(s)}$, and 3-prong.
In the \Htt channel, events with jets originating from \cPqb\ quark decays are excluded in order to reduce the background from $\ttbar$ events.
The combined secondary vertex algorithm~\cite{csv3} is used to
identify jets originating from \cPqb\ quarks by their characteristic displaced vertices.

The missing transverse momentum vector (\ptvecmiss), with magnitude \ptmiss, is the negative vector sum of the \pt of all PF candidates in an event.
Jet energy corrections are propagated to the \ptvecmiss for a more accurate measurement~\cite{CMS_MET}.
Events may have anomalously large \ptmiss from sources such as detector noise, cosmic ray muons, and beam halo particles, which are not well modeled in simulation.
Event filters~\cite{CMS_MET_PAS} are applied to remove such events.

\section{Observed and simulated data samples \label{sec:Samples}}

The analysis is performed with $\Pp\Pp$ collision data at $\sqrt{s} = 13 \TeV$ collected with the CMS detector in 2016.
The data correspond to an integrated luminosity of 35.9\fbinv.

The analysis strategy and event selection were optimized using Monte Carlo (MC) simulated samples of associated DM+\Ph production
via the two benchmark models discussed in Section~\ref{sec:Introduction}.
The \MGvATNLO~v2.3.0~\cite{aMCatNLO} generator is used to generate both the \PZpr-2HDM and baryonic \PZpr signals at LO.
The decay of the SM-like Higgs boson is simulated by \PYTHIA~8.205~\cite{Sjostrand:2014zea}.

A small but irreducible background for both decay channels in this analysis
comes from events in which the SM Higgs boson is produced in association with a \PZ boson that decays to two neutrinos.
Other SM Higgs boson production mechanisms are associated with resonant but reducible backgrounds.
These include gluon-gluon fusion ($\cPg\cPg\Ph$), vector boson fusion (VBF), and production in association with a pair of top quarks ($\ttbar$\Ph).
The production in association with a vector boson (V\Ph) and other SM Higgs boson backgrounds are all generated using \MGvATNLO~v2.2.2
at next-to-leading order (NLO) in perturbative QCD.

The dominant nonresonant backgrounds for the \Hgg channel are events with mismeasured \ptmiss and two photons
that happen to have an invariant mass close to the mass of the SM Higgs boson.
The largest contributions to this are nonresonant $\gamma\gamma$, \GAMJET\xspace, and QCD multijet production.
The simulated $\gamma\gamma$ sample is generated at LO with \SHERPA~v2.2.2~\cite{sherpa}
while the \GAMJET\xspace and QCD multijet samples are modeled at LO with \PYTHIA.
Additional backgrounds originate from electroweak (EW) processes such as single top, $\ttbar$, \PW, or \PZ boson
production in association with one or two photons, and
Drell--Yan (DY) production where the \PZ boson decays to pairs of electrons or muons.
The DY and all other EW backgrounds considered in the analysis are generated at NLO with \MGvATNLO.
These nonresonant background samples are used for optimizing the analysis selection,
however, they are not used for the ultimate background estimation.

The largest backgrounds for the \Htt channel are \wjet, $\ttbar$, and multiboson processes.
The \MGvATNLO~v2.3.0 generator is used for \wjet processes,
which are generated at LO in perturbative QCD with the MLM jet matching and merging scheme~\cite{Alwall:2007fs}.
A $\pt$-dependent correction factor is applied to the \wjet sample to account for next-to-next-to-leading order QCD and NLO EW effects~\cite{Sirunyan:2017hci,Kuhn:2005gv,Kallweit:2014xda,Kallweit:2015dum}.
The $\ttbar$ process is generated at NLO with the \POWHEG 2.0~\cite{Frixione:2007vw, Alioli:2010xd, Alioli:2010xa, Alioli:2008tz} generator.
Single top quark production is modeled at NLO with the \POWHEG 1.0~\cite{powheg1} generator.
The FxFx~\cite{FXFX} merging scheme is used to generate some smaller diboson backgrounds (including $\PW\PZ$ samples) with the \MGvATNLO generator at NLO, while the dominant diboson backgrounds, $\PW\PW$ and $\PZ\PZ$ in two lepton final states, are generated using \POWHEG 2.0.
Another reducible background considered in this analysis is $\cPZ/\gamma^* \to \ell\ell/\Pgt\Pgt$, where $\ell$ is $\Pe$ or $\Pgm$.
The Drell--Yan background is corrected for differences in the dilepton mass $m_{\ell\ell/\Pgt\Pgt}$ and dilepton transverse momentum $\pt(\ell\ell/\Pgt\Pgt)$ distributions using dimuon events in data~\cite{Sirunyan:2017khh}.

All simulated samples mentioned above use the NNPDF 3.0 parton distribution function (PDF) sets~\cite{Ball:2014uwa,Skands:2014pea}
with the order matching that used in the matrix element calculations.
For parton showering and hadronization, as well as for $\tau$ lepton decays, the samples are interfaced with \PYTHIA using the CUETP8M1 tune~\cite{Khachatryan:2015pea} for all samples except \ttbar, for which the M2 tune is used.
The MC samples are processed through a full simulation of the CMS detector based on \GEANTfour~\cite{Agostinelli:2002hh}
and are reconstructed with the same algorithms that are used for the data.
All samples include the simulation of additional inelastic $\Pp\Pp$ interactions in the same or neighboring bunch crossings (pileup).
Minimum-bias collision events generated with \PYTHIA are added to the simulated samples to reproduce the pileup effects in the data.
Additionally, the simulated events are weighted so that the pileup vertex distribution matches
that of the data, with an average of 27 interactions per bunch crossing.

\section{Analysis strategy in the \texorpdfstring{\Hgg}{h to gg} channel \label{sec:AnalysisHgg}}

The search for DM+\Ph in the \Hgg channel is performed by selecting events with two photons and a large amount of \ptmiss.
The set of requirements detailed in Section~\ref{sec:HggSelection} is applied to select well-identified photons and to enhance the signal significance.
A fit to the diphoton invariant mass distribution, described in Section~\ref{sec:HggExtraction}, is performed to extract the background and signal yields.

\subsection{Event selection \label{sec:HggSelection} }

The events used in this analysis were selected by a diphoton trigger with asymmetric $\pt$ thresholds of 30 and 18\GeV and diphoton invariant mass above 90\GeV.
The trigger also has loose photon identification criteria based on the cluster shower shape, isolation requirements,
and a selection on the ratio of hadronic to electromagnetic energy deposits of the photon candidates.

The photons that enter the analysis are required to fall within the fiducial range of the ECAL ($\abs{\eta} < 1.44$ or $1.57 < \abs{\eta} < 2.50$)
and to satisfy various preselection criteria that are slightly more stringent than the trigger requirements.
An additional veto on the presence of a track pointing to the ECAL cluster is applied to reject electrons that could be reconstructed as photons.
Scale factors, extracted from $\PZ\to\Pe^+\Pe^-$ events using the tag-and-probe method~\cite{WZFirstPaper},
are applied to the simulated samples to account for any discrepancy in identification efficiency between data and simulation.

The isolation variables that are used in the photon identification are calculated by summing
the \pt of PF photons, neutral hadrons, or charged hadrons associated with the PV
in a cone of radius $\Delta R = \sqrt{\smash[b]{(\Delta\eta)^2+(\Delta\phi)^2}} = 0.3$.
The isolation variables are corrected by the median transverse momentum density of the event
to mitigate the effects of pileup~\cite{pileupSubtract}.
Some of the signals considered can have Lorentz-boosted topologies.
For example, high-mass mediators could result in a large boost to the Higgs boson.
When a boosted Higgs boson decays to two photons, the resulting photons hit the ECAL close to each other.
This effect leads to large contributions from one photon to the photon isolation sum of the other.
In order to maintain high efficiency for high-mass mediator signals,
the photon isolation requirement is not applied to photons that are within $\Delta R < 0.3$ of each other.

Preselected photons are required to have leading (subleading) photon \pt above 30 (20)\GeV and diphoton invariant mass \mgg above 95\GeV.
Simulated signal and background samples that pass the preselection were used to study the discriminating power of variables such as
\ptmiss, the \pt of the diphoton system \ptgg, and the ratio \pt/\mgg for each photon.
A selection on \pt that scales with \mgg is chosen so that it does not distort the shape of the \mgg distribution.
The \ptgg variable is included in the selection because it has higher resolution than the event's measured \ptmiss and is expected to be large
for signal events, since the Higgs boson is produced back-to-back with \ptvecmiss.
A high-\ptmiss category ($\ptmiss \ge 130 \GeV$) is optimal for the two benchmark models presented in this paper.
A low-\ptmiss category ($50 < \ptmiss < 130 \GeV$), optimized using as reference the baryonic \PZpr signal model, is also included to probe softer signals,
namely signals that may not be observed in other $\Ph+\ptmiss$ searches because they rely heavily on \ptmiss for background rejection.
The chosen requirements, found to optimize the signal sensitivity for both models in the low- and high-\ptmiss categories, are given in Table~\ref{tab:HggSelection}.

\begin{table}[h!t]
  \centering
  \topcaption{ Optimized kinematic requirements for the low- and high-\ptmiss categories. \label{tab:HggSelection} }
  \begin{tabular}{lcc}
  Variable                           & Low-\ptmiss category     & High-\ptmiss category \\
  \hline
  \ptmiss                            & ${>}50\GeV$, ${<}130\GeV$	& ${>}130\GeV$ \\
  $p_{\mathrm{T}1}/\mgg$ 	     & ${>}0.45$         		& ${>}0.5$\\
  $p_{\mathrm{T}2}/\mgg$ 	     & ${>}0.25$         		& ${>}0.25$\\
  \ptgg                              & ${>}75\GeV$    		& ${>}90\GeV$ \\
  \end{tabular}
\end{table}

\begin{figure}
  \centering
  \includegraphics[width=0.6\textwidth]{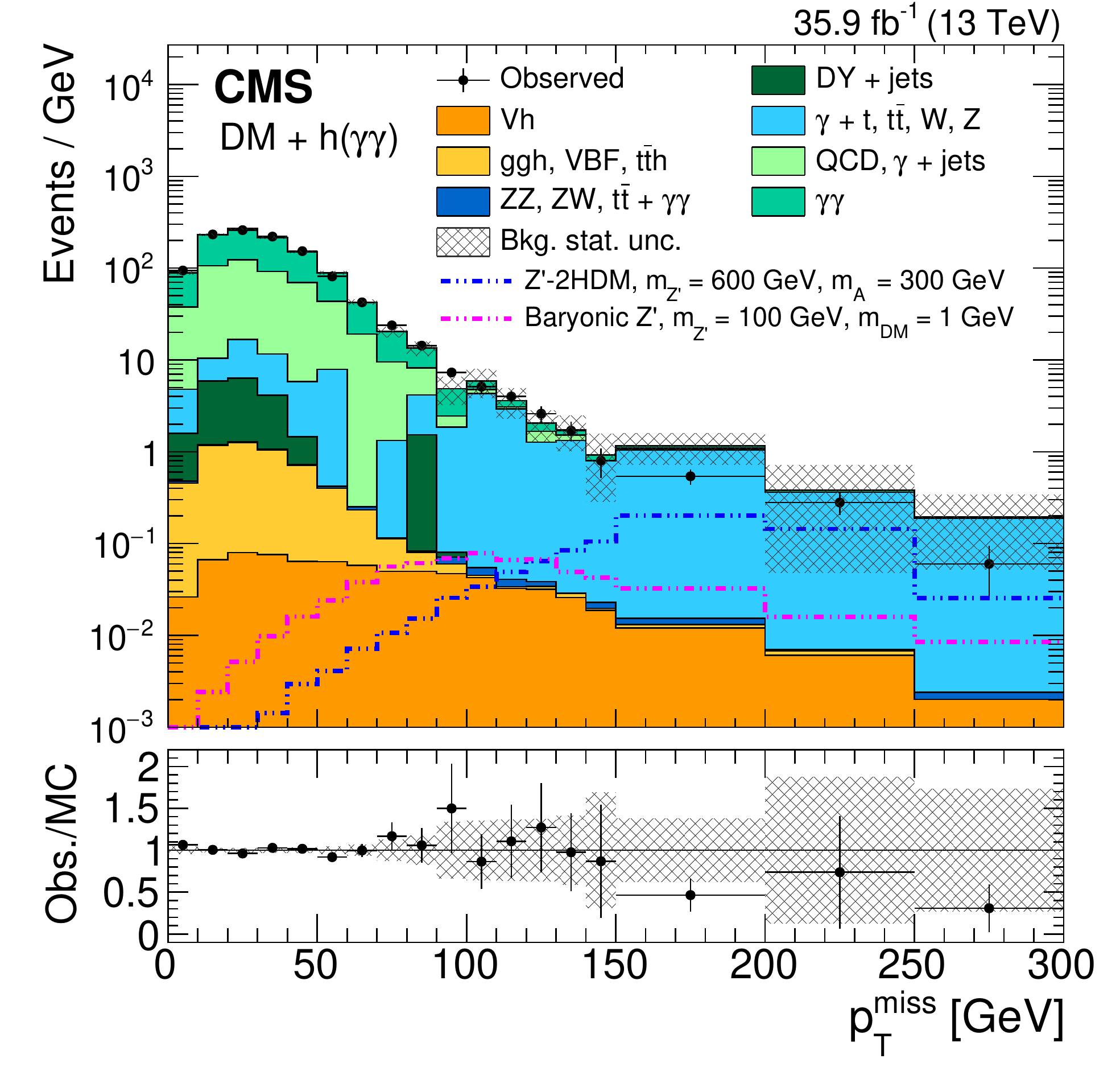}
  \caption{\label{fig:HggMET}
Distribution of \ptmiss for events passing the requirements given in Table~\ref{tab:HggSelection}.
Events with \ptmiss below 50\GeV are not used in the analysis.
The cross sections of the signals are set to 1\unit{pb}.
The total simulated background is normalized to the integral of the data.
The statistical uncertainty in the total background is shown by the hatched bands.
The data-to-simulation ratio is shown in the lower panel.
}
\end{figure}

Further background rejection is achieved using two topological requirements.
The azimuthal separation $\abs{\Delta\phi(p_{\gamma\gamma},\ptvecmiss)}$ between \ptvecmiss and the Higgs boson direction
reconstructed from the two photons must be greater than 2.1 to select events in which the Higgs boson and \ptvecmiss are back-to-back.
Events with highly energetic jets collinear to \ptvecmiss are removed by the requirement
that the $\min\abs{\Delta\phi(p_\text{jet},\ptvecmiss)}$ be greater than 0.5 for any jet with \pt above 50\GeV.
This rejects events with a large misreconstructed \ptmiss arising from mismeasured jet \pt.
Finally, events are vetoed if they have three or more jets each with \pt above 30\GeV, to reject multijet backgrounds
while maintaining a high efficiency for the two benchmark signal models.
The \ptmiss distribution of the selected events is shown in Fig.~\ref{fig:HggMET}.

\subsection{Background estimation and signal extraction \label{sec:HggExtraction} }

A narrow resonance search similar to the SM Higgs boson diphoton analysis of Ref.~\cite{smhgg} is performed.
The diphoton invariant mass between 105 and 180\GeV is fit with a model that is the sum of the signal and background shapes.
The signal shape, taken from the simulated events, is allowed to change independently in each of the two \ptmiss categories.
The background shape includes a smooth function, estimated from the data,
to model the continuum background, and a resonant contribution from the SM Higgs boson.
The fit is performed with an unbinned maximum-likelihood technique in both the low- and high-\ptmiss categories discussed above.

The resonant background, arising from the SM Higgs boson decays to two photons, appears as a peak under the expected signal peak.
This contribution from all SM Higgs boson production modes is estimated with the simulated events by including a mass distribution template,
scaled to the NLO cross section, as a resonant component in the final fitting probability density function (pdf).

The nonresonant background contribution, mostly due to $\gamma\gamma$ and to various EW processes,
is estimated using data.
The nonresonant diphoton \mgg distribution in the data is fit, in each \ptmiss category, with an analytic function.
Because the exact functional form of the background is unknown, the parametric
model must be flexible enough to describe a variety of potential underlying functions.
Using an incorrect background model can lead to biases in the measured signal yield that
can artificially modify the sensitivity of the analysis.
Three functions are considered as possible models for the nonresonant background; they are analytical forms
that are frequently used in dijet~\cite{dijet} and diphoton~\cite{highmassdipho} resonance searches.
The best functional form found to fit the nonresonant diphoton \mgg distribution, in both \ptmiss categories,
is a power law function $f(x) = ax^{-b}$ where $a$ and $b$ are free parameters constrained to be positive.

A detailed bias study has been performed in order to choose this function.
The \mgg shape of the simulated nonresonant events is used as a template to generate 1000 pseudo-experiments for each \ptmiss category.
For each pseudo-experiment, the number of events generated is equal to the number of events observed in data in that category.
The resulting \mgg distribution is fit with each analytic function considered.
The exercise is also repeated injecting a potential signal contribution.
The pulls of each pseudo-experiment, defined as the difference in the number of simulated events and those predicted by the fit function
divided by the statistical uncertainties of the fit, are calculated.
If the bias (the median of the pulls) is five times smaller than the statistical uncertainty
in the number of fitted signal events, any potential bias from the choice of background model is considered negligible.
Since this criterion is satisfied for the power law function, any systematic uncertainty in the bias from the background fit function
is neglected in this analysis.

\begin{figure}
  \centering
  \includegraphics[width=0.49\textwidth]{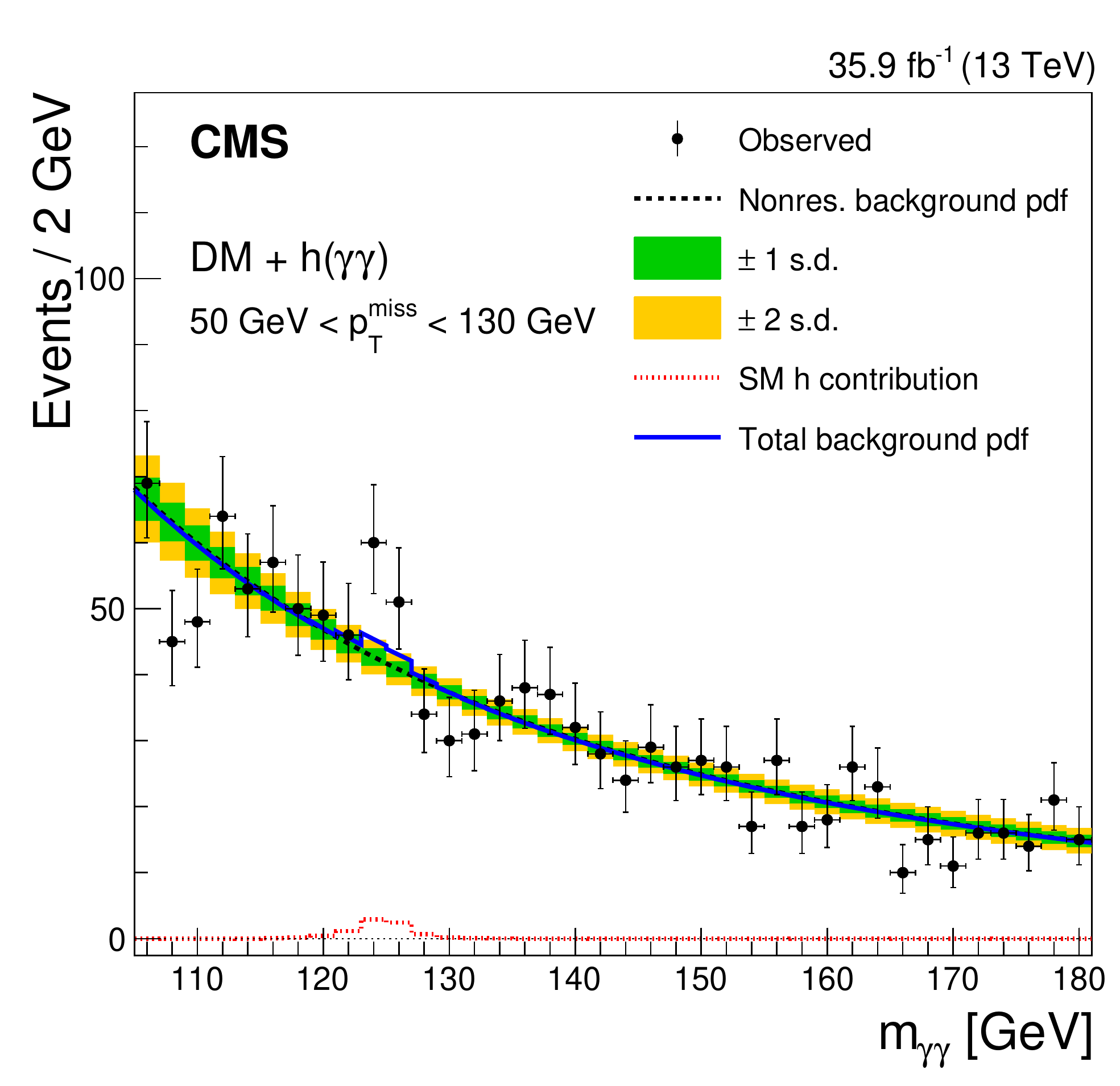}
  \includegraphics[width=0.49\textwidth]{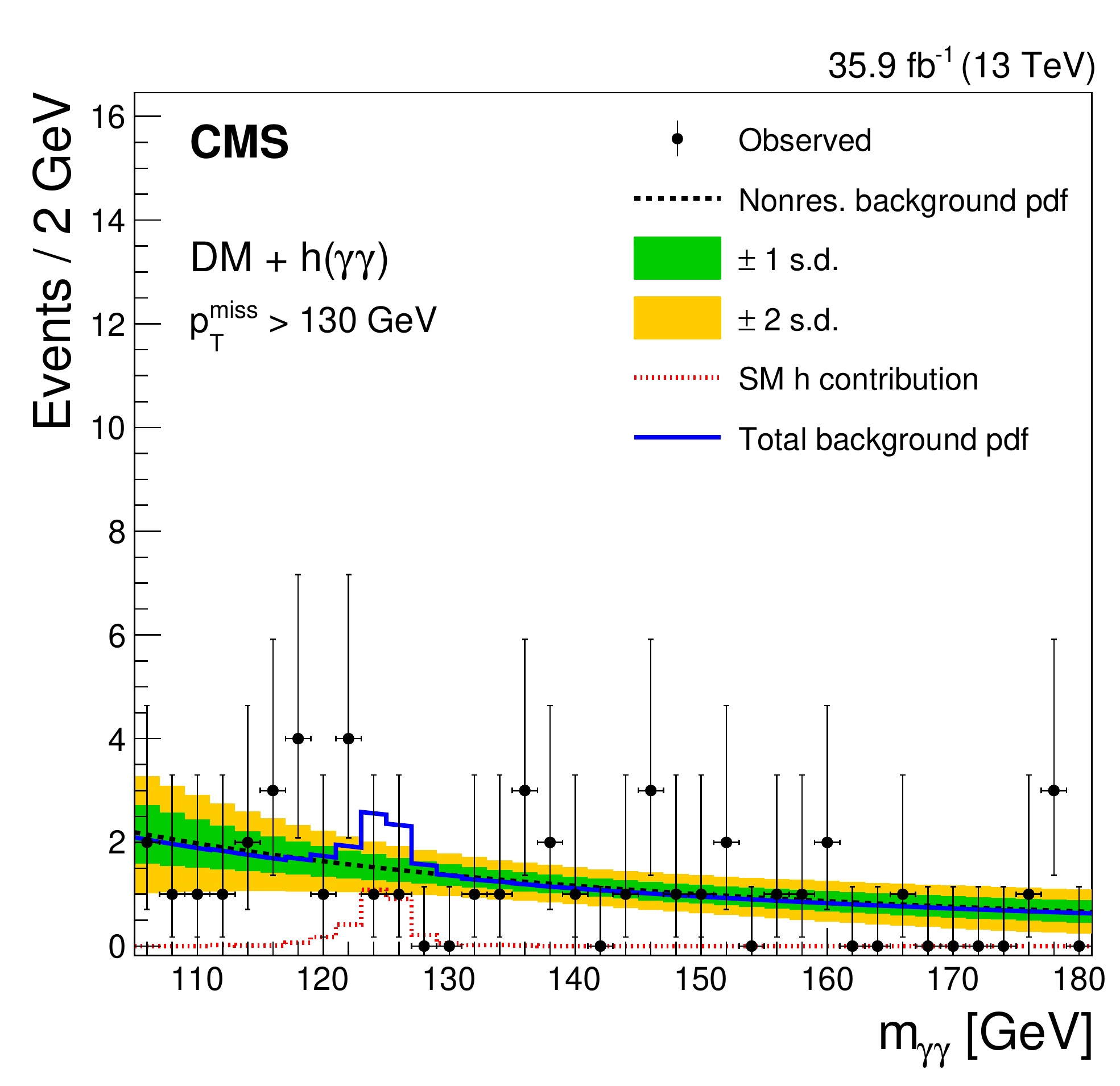}
  \caption{\label{fig:HggFit}
The background-only fit to data is performed, for low-\ptmiss (left) and high-\ptmiss (right) categories,
with the sum of a power law (dashed black) fit function to describe the nonresonant contribution,
and a resonant shape (dashed red), taken from simulation, to take into account the SM \Hgg contribution.
The SM \Ph contribution is fixed to the theoretical prediction in the statistical analysis.
The sum of the nonresonant and resonant shapes (solid blue) is used to estimate the total background in this analysis.
  }
\end{figure}

The final background-only fit for both \ptmiss categories is shown in Fig.~\ref{fig:HggFit}.
Both the resonant and nonresonant background pdf contributions are shown.
The slight excess of events observed in data around 125\GeV in the low-\ptmiss category
is compatible with the SM Higgs boson expectation within 2.0 standard deviations.

\section{Analysis strategy in the \texorpdfstring{\Htt}{h to tautau} channel \label{sec:AnalysisHtt}}

\subsection{Event selection}

The three final states of $\Pgt$ lepton pairs with the highest $\Pgt\Pgt$ branching fractions ($\Pe\tauh$, $\Pgm\tauh$, and $\tauh\tauh$) are considered in this analysis.
In the $\Pe\tauh$ and $\Pgm\tauh$ channels, one of the $\Pgt$ leptons decays leptonically to an electron or a muon and two neutrinos, while the other $\Pgt$ lepton decays hadronically ($\tauh$) with one neutrino. In the third channel, $\tauh\tauh$, both $\Pgt$ leptons decay hadronically.
The $\Pe\Pgm$, $\Pe\Pe$, and $\Pgm\Pgm$ final states are not included because of the low branching fraction of the $\Pgt\Pgt$ pair to purely leptonic final states. The $\Pe\Pe$ and $\Pgm\Pgm$ final states are not considered, since they are overwhelmed by DY background.

Triggers based on the presence of a single electron (muon) are used to select events in the $\Pe\tauh$ ($\Pgm\tauh$) channel. In the $\tauh\tauh$ channel, the triggers require the presence of two isolated $\tauh$ objects. Each $\tauh$ candidate reconstructed offline is required to match a $\tauh$ candidate at the trigger level, with a $\Delta R$ separation less than 0.5.

The electrons and muons in the $\Pe\tauh$ and $\Pgm\tauh$ channels are required to have $\pt$ greater than $26\GeV$, exceeding the trigger thresholds for the single-electron and single-muon triggers.
Electrons (muons) with $\abs{\eta} < 2.1$ (2.4) are used.
An $\Pe\tauh$ ($\Pgm\tauh$) event is required to have an electron (muon) passing a multivariate MVA identification discriminator~\cite{hpssource}
and an isolation requirement of $I^{\Pe}_{\text{rel}} < 0.10$ ($I^{\Pgm}_{\text{rel}}< 0.15$),
where $I^\ell_{\text{rel}}$ is defined as in Eq.~(\ref{eqn:lepiso}), with an isolation cone of size $\Delta R=0.3$ $(0.4)$ surrounding the electron (muon):
\begin{equation}
     I^{\ell}_{\text{rel}} =  \left(\sum \pt^{\text{charged}}+\max[0, \sum \pt^{\text{neutral}} + \sum \pt^{\gamma} - 0.5 \times \sum \pt^{\mathrm{{PU}}}]\right)/\pt^{\ell}.
     \label{eqn:lepiso}
 \end{equation}
 Here $\sum \pt^{\text{charged}}$, $\sum \pt^{\text{neutral}}$, and $\sum \pt^{\gamma}$ are the scalar sums of transverse momentum from charged hadrons associated with the primary vertex, neutral hadrons, and photons, respectively.
The term $\sum \pt^{\mathrm{PU}}$ is the sum of transverse momentum of charged hadrons not associated with the primary vertex and $\pt^{\ell}$ is the $\pt$ of the electron or muon.

Hadronically decaying $\tau$ leptons in all channels are required to satisfy a loose ($\tauh\tauh$ channel)
or a tight ($\Pe\tauh$ and $\Pgm\tauh$ channels) working point of an MVA isolation measure.
The loose (tight) working point corresponds to a 65 (50)\% efficiency with a 0.8 (0.2)\% misidentification probability.
The $\tau$ leptons are required to be identified as decaying via one of the three modes recognized with the HPS algorithm,
and also pass discriminators that reduce the rate of electrons and muons misreconstructed as $\tauh$ candidates~\cite{tauidsource}.
For the $\Pe\tauh$ and $\Pgm\tauh$ channels, the $\tauh$ candidates are required to have $\pt>20\GeV$ and $\abs{\eta}<2.3$.
In the $\tauh\tauh$ channel, the leading (subleading) $\tau$ lepton \pt is required to be greater than 55 (40)\GeV,
both $\tauh$ transverse momenta exceeding the double-hadronic $\tau$ lepton trigger thresholds of 35\GeV.
The selection criteria are summarized in Table~\ref{tab:inclusive_selection} for all three final states.

 \begin{table}[htbp]
     \centering
     \topcaption{Selection requirements for the three $\Pgt\Pgt$ decay channels.
         The \pt thresholds for the triggers are given in the second column in parentheses.
         \label{tab:inclusive_selection}
     }

     \begin{tabular}{ccccc}
                       &  &  \multicolumn{3}{c}{Lepton selection}  \\
         Final state   &  Trigger type  & \pt $[{\GeVns}]$    & $\eta$    &  Isolation \\
         \hline
         $\Pe\tauh$    &  \Pe (25\GeV)                    &  $\pt^\Pe>26$                          & $\abs{\eta^\Pe}<2.1$      &  $I^{\Pe}_{\text{rel}} <0.1$    \\
                       &                                &  $\pt^{\tauh}>20$                      &  $\abs{\eta^{\tauh}}<2.3$ &  Tight MVA $\tauh$    \\
         $\Pgm\tauh$    &  \Pgm (24\GeV)                    &  $\pt^\Pgm>26$                         &  $\abs{\eta^\Pgm}<2.4$    &  $I^{\Pgm}_{\text{rel}} <0.15$  \\
                       &                                &  $\pt^{\tauh}>20$                      &  $\abs{\eta^{\tauh}}<2.3$ &  Tight MVA $\tauh$    \\
         $\tauh\tauh$  &  \tauh (35\GeV) \& \tauh (35\GeV)  &  $\pt^{\tauh}>55\,\&\,40$              & $\abs{\eta^{\tauh}}<2.1$  &  Loose MVA $\tauh$    \\
     \end{tabular}

 \end{table}

The \ptmiss is further required to be greater than 105\GeV and the visible \pt of the $\Pgt\Pgt$ system is required to be greater than 65\GeV.
These stringent criteria reduce the need for tighter isolation in the $\tauh\tauh$ channel.
Additionally, the mass reconstructed from the visible \pt of the $\Pgt\Pgt$ system is required to be less than 125\GeV,
to ensure that the $\Pgt\Pgt$ system is compatible with an SM Higgs boson.
In order to minimize diboson and \wjet contributions, the two $\Pgt$ lepton candidates must pass a loose collinearity criterion of $\Delta R_{\Pgt\Pgt}<2.0$.

Two types of event veto are employed for background reduction.
Events with jets tagged as originating from hadronization of \cPqb\ quarks are vetoed, to reduce \ttbar and single top processes.
The working point used in the \cPqb\ tagging algorithm corresponds to about a 66\% efficiency for a 1\% misidentification probability.
In addition, events with additional muons or electrons beyond those from the $\Pgt$ lepton candidates are discarded,
to reduce the contribution of multilepton backgrounds.

\subsection{Signal extraction and background estimation}

The signal is extracted from a maximum-likelihood fit to the total transverse mass (\mttot) distributions in the different channels for the SR, and for the \wjet and QCD multijet background CRs. The \mttot is defined as:
\begin{equation}
    \mttot = \sqrt{\smash[b]{
        (p_\mathrm{T}^{\Pgt_1}+p_\mathrm{T}^{\Pgt_2}+\ptmiss)^2
        - (p_x^{\Pgt_1}+p_x^{\Pgt_2}+p_x^{\text{miss}})^2
        -(p_y^{\Pgt_1}+p_y^{\Pgt_2}+p_y^{\text{miss}})^2
    }},
    \label{eqn:mttot}
\end{equation}
where $p_{x}^{\text{miss}}$ and $p_{y}^{\text{miss}}$ are the magnitudes of the $x$ and $y$ components of \ptvecmiss, respectively.

The \wjet and the QCD multijet background are estimated directly from the data.
The procedure to estimate these processes relies on CRs, which are included in the maximum-likelihood fit, to extract the results.
The other backgrounds, $\ttbar$, \zjet, SM Higgs boson, single top quark, and diboson production processes, are extracted from simulation.

The shape of the \mttot distribution of the \wjet background is estimated from simulation by requiring the same selection as for the SR,
but the isolation of the $\Pgt$ lepton candidates is relaxed to increase the statistical precision of the distribution.
To constrain the normalization of the \wjet background, a CR enriched in \wjet events is constructed by inverting
the isolation criteria on the \tauh candidates while keeping a loose isolation.
The CR obtained by inverting the isolation criterion is included in the maximum-likelihood fit to constrain
the normalization of the \wjet background in the SR.

To estimate the QCD multijet background, a CR in data is obtained by requiring the $\Pgt$ lepton candidates to have the same sign.
No significant amount of signal and of background with opposite-sign \tauh is expected in this CR
because the \tauh charge misidentification is of order 1\% and the charge misidentification for electrons and muons is even smaller.
All simulated backgrounds are subtracted from observed events in the CR, and the remaining contribution is classified as QCD multijet background.
The contribution of QCD multijet events with opposite-sign $\Pgt$ lepton candidates in the SR is obtained by multiplying
the QCD multijet background, obtained in the same-sign CR, by a scale factor.
The scale factor, approximately unity with an uncertainty of ~20\%, is determined from events with \tauh candidates
failing the isolation requirement and with low \ptmiss, which do not overlap with events selected in the SR.
To increase the statistical precision of the QCD multijet distribution, the isolation of the $\Pgt$ lepton candidates
is relaxed for the $\ell\tauh$ channels, while conserving the normalization obtained as detailed above.
The same-sign $\Pgt$ lepton candidate CR constrains the QCD multijet background normalization in the SR and the other CRs in the maximum-likelihood fit.

\begin{figure}[htbp]
    \centering
    \includegraphics[width=0.49\textwidth]{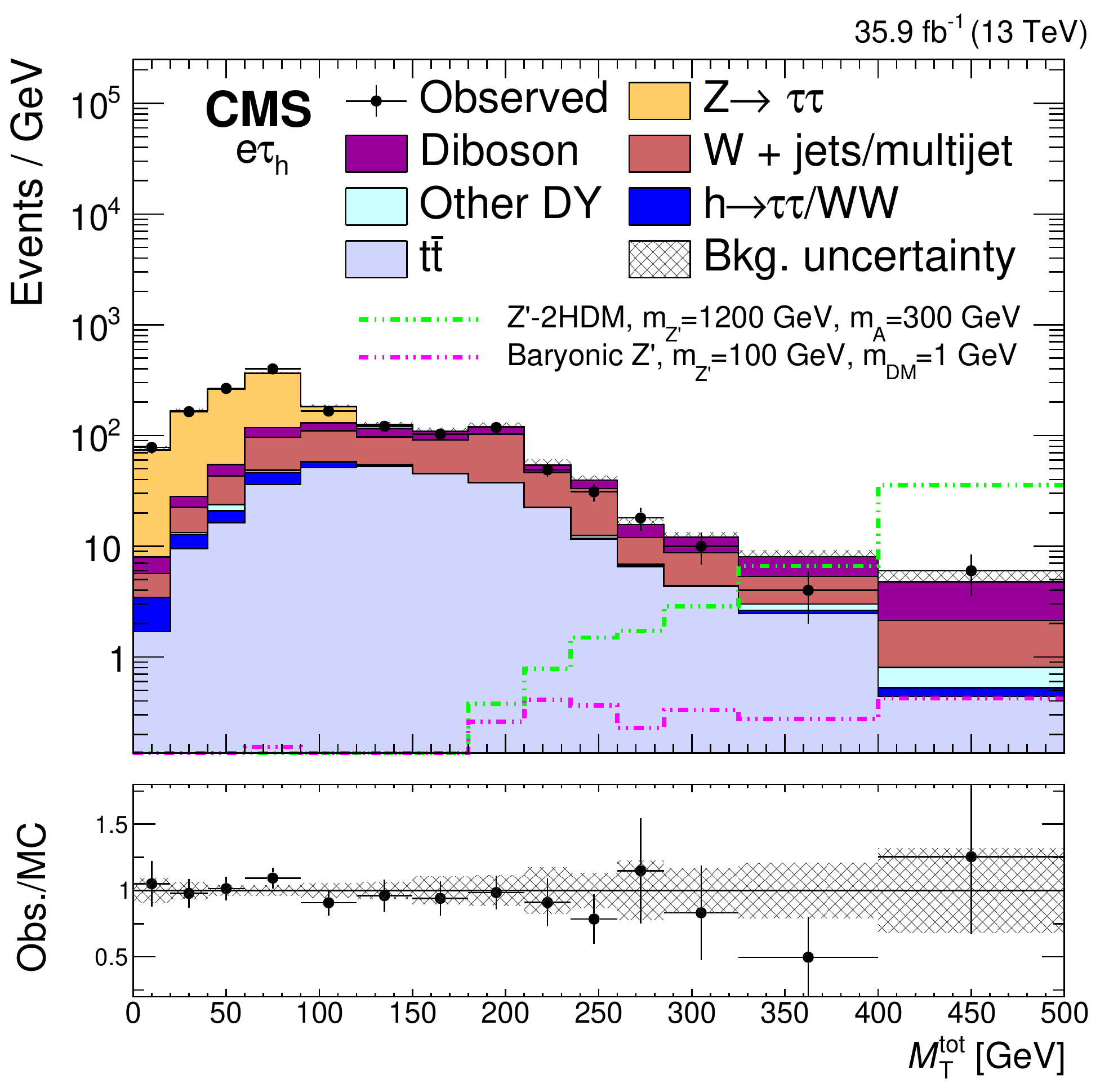}
    \includegraphics[width=0.49\textwidth]{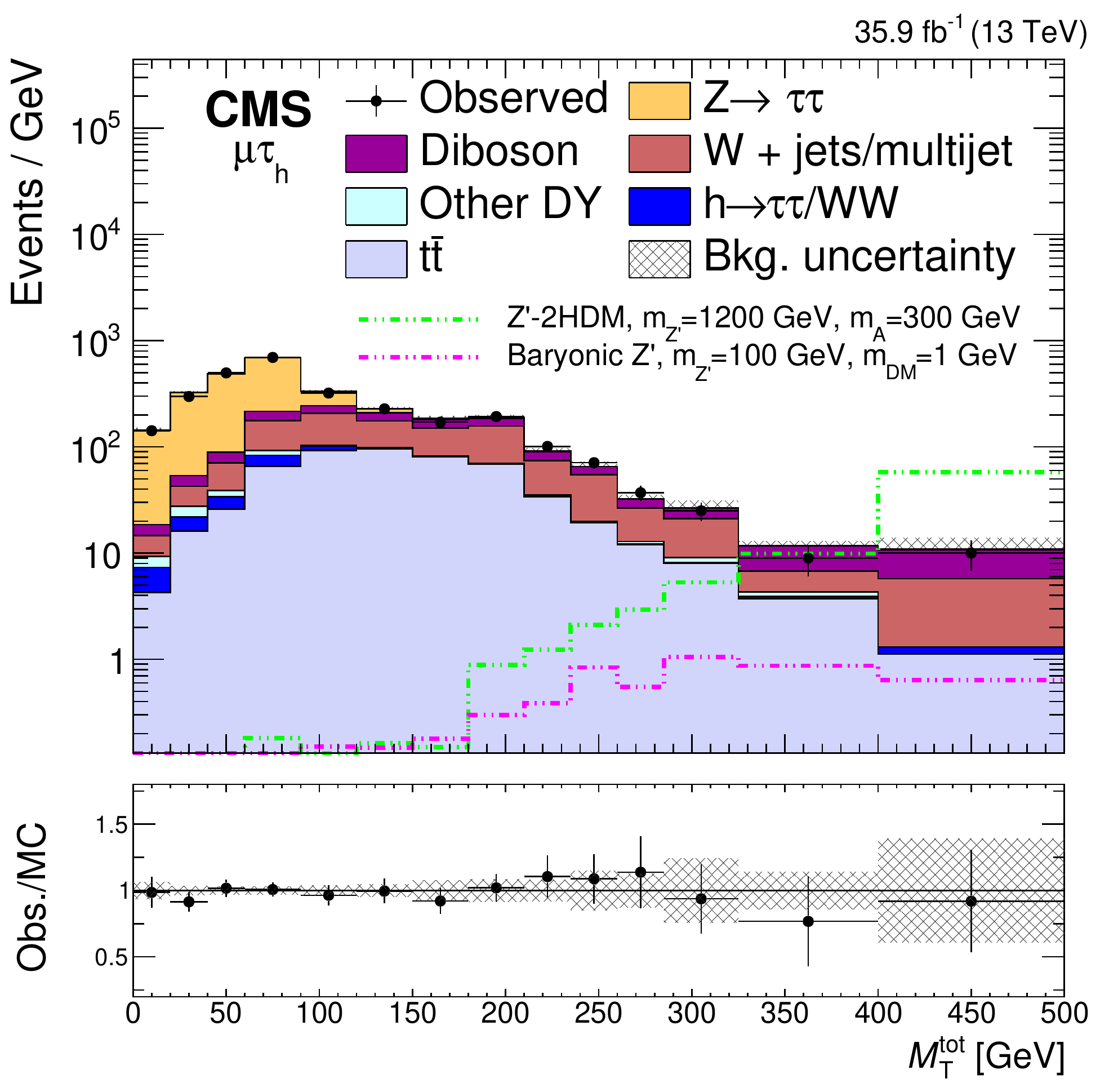}\\
    \includegraphics[width=0.49\textwidth]{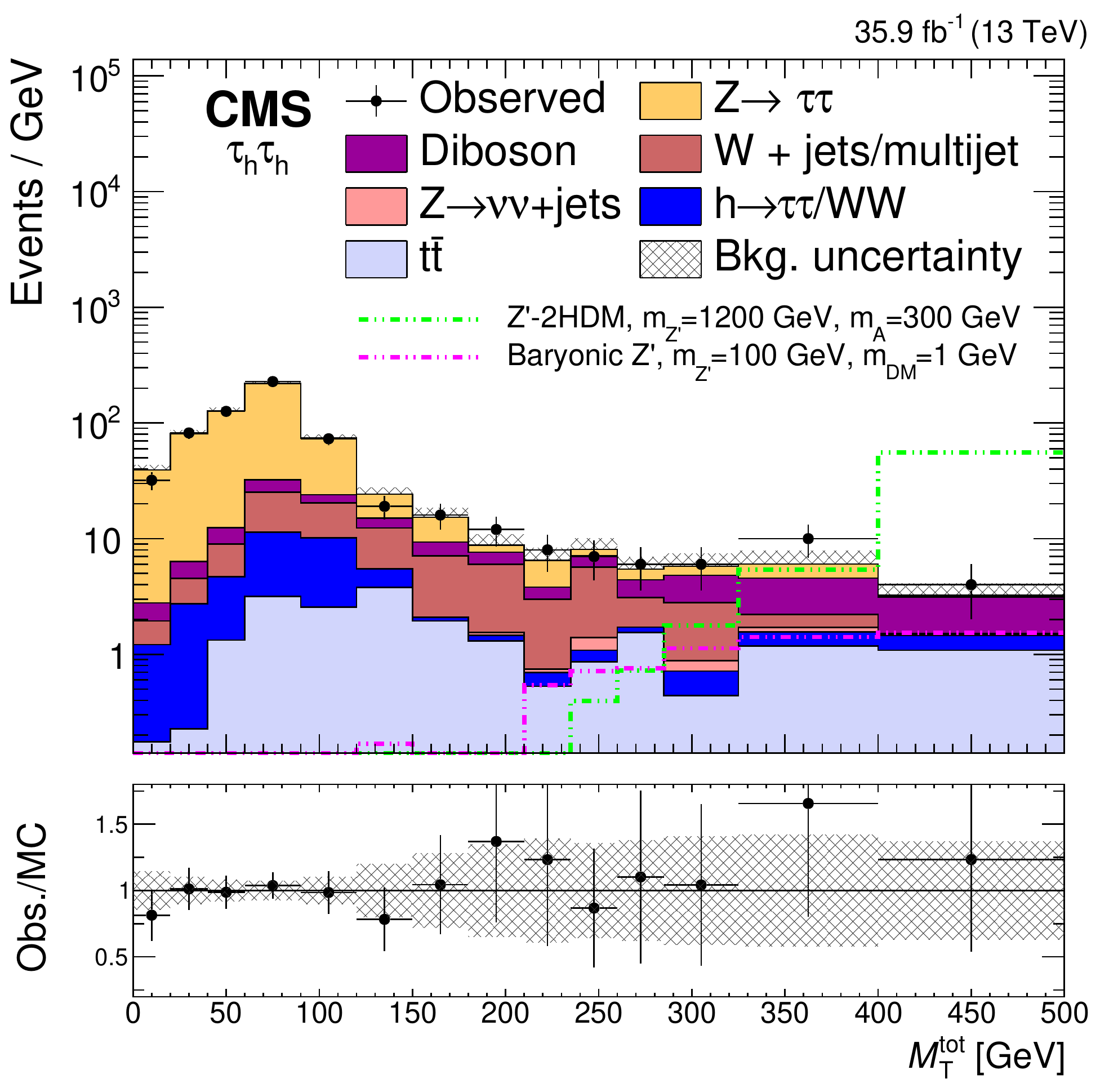}
    \caption{
        Distributions of the total transverse mass \mttot in the SR for the $\Pe\tauh$ (upper left), $\Pgm\tauh$ (upper right), and $\tauh\tauh$ (lower) final states are shown after the simultaneous maximum-likelihood fit.
        Representative signal distributions are shown with cross sections normalized to 1\unit{pb}.
        The data points are shown with their statistical uncertainties, and the point in the final bin includes overflow.
        The statistical uncertainty of the observed distribution is represented by the error bars on the data points.
        The overflow of each distribution is included in the final 400--500\GeV bin.
        Single top processes are included in the ``Diboson'' contribution. The ``Other DY'' contribution includes background from $\PZ\to\ell\ell$.
        The systematic uncertainty related to the background prediction is indicated by the shaded band.
    }
    \label{fig:postfit_mt}
\end{figure}

The normalizations of the \wjet and QCD multijet background are strongly correlated since both processes contribute to both CRs.
The simultaneous fit of the SR and CRs takes into account this correlation.
The SR distributions included in the simultaneous maximum-likelihood fit are shown in Fig.~\ref{fig:postfit_mt}.
For the signal extraction, \wjet and QCD multijet background CRs are considered separately, whereas in
Fig.~\ref{fig:postfit_mt}, the two backgrounds are presented merged together.

\section{Systematic uncertainties \label{sec:Systematics}}

In both analysis channels, an uncertainty of 2.5\% is used for the normalization of simulated samples to reflect
the uncertainty in the integrated luminosity measurement in 2016~\cite{CMS-PAS-LUM-17-001}.
Common to both analysis channels are systematic uncertainties related to the
theoretical production cross section of the Higgs boson.
The PDF, and renormalization and factorization scale uncertainties
are addressed using the recommendations of PDF4LHC~\cite{systematics_PDF4LHC} and LHC Higgs Cross Section~\cite{deFlorian:2016spz} working groups, respectively.
The value of these uncertainties range from 0.3 to 9.0\%.
The systematic uncertainties associated with each of the analysis channels are detailed below.
Uncertainties affecting normalizations are represented by log-normal pdfs in the statistical analysis.

\subsection{The \texorpdfstring{\Hgg}{h to gg} channel}

In the \Hgg channel,
there are several sources of experimental and theoretical uncertainties that affect the signal and the SM \Hgg yields.
However, the largest source of uncertainty is statistical.
As mentioned in Section~\ref{sec:HggExtraction}, no systematic uncertainties are applied to the nonresonant background, which is extracted from a fit to data,
since the bias of the fit is negligible compared to the statistical uncertainty of the data set.
The systematic uncertainties for the \Hgg channel are summarized in Table~\ref{tab:HggSystematics}.
In addition to the theoretical uncertainties mentioned above, a 20\% cross section uncertainty is included for the $\Pg\Pg\Ph$ sample,
based on the CMS differential measurements of \Hgg, for diphoton \pt above 70\GeV~\cite{diffXsecHgg}.
The branching fraction uncertainty~\cite{deFlorian:2016spz} of 1.73\% is also included.

In addition to the integrated luminosity uncertainty, several other experimental sources of systematic uncertainty are included in this analysis.
The trigger efficiency uncertainty (approximately 1\%) is extracted from  $\PZ\to\Pe^+\Pe^-$ events using a tag-and-probe technique~\cite{WZFirstPaper}.
The photon identification uncertainty of 2\% arises from the observed difference in efficiencies between data and simulation.
A 0.5\% energy scale uncertainty is assigned to take into account the knowledge of the photon energy scale at the \PZ boson mass peak and its extrapolation to the Higgs boson mass.
Additionally, several \ptmiss-related uncertainties are applied.
The systematic uncertainty from mismeasured \ptmiss is evaluated by comparing the tail of the \ptmiss distributions in data and simulation in a \GAMJET\xspace enriched CR.
The efficiencies with which data and simulated events pass the \ptmiss selection are compared.
The difference in efficiency is 50\% and is included as a systematic uncertainty associated with mismeasured \ptmiss.
However, the contribution of simulated backgrounds with mismeasured \ptmiss is quite small
since only the $\Pg\Pg\Ph$ and VBF SM \Hgg production modes contribute.
Finally, a systematic uncertainty, which is less than 4\%, is applied to take into account the difference in efficiency between
data and simulation when applying the topological $\Delta\phi$ requirements in the low-\ptmiss region.
This uncertainty is evaluated using $\PZ\to \Pe^+\Pe^-$ events, and only affects the $\Pg\Pg\Ph$ and VBF simulated samples.

\begin{table}[ht!]
\centering
\topcaption{
Systematic uncertainties affecting the signal and resonant backgrounds in the \Hgg channel.
\label{tab:HggSystematics}}
\begin{tabular}{lcc}
& Signal [\%] & SM \Ph [\%] \\
\hline
Theoretical sources & & \\
\hspace*{5mm} PDF & \NA & 2--4 \\
\hspace*{5mm} Renorm. and fact. scale & \NA & 0.3--9 \\
\hspace*{5mm} Cross section ($\Pg\Pg\Ph$) & \NA & 20 \\
\hspace*{5mm} Higgs boson branching fraction & \multicolumn{2}{c}{ 1.73 } \\
Experimental sources & & \\
\hspace*{5mm} Integrated luminosity & \multicolumn{2}{c}{ 2.5 } \\
\hspace*{5mm} Trigger efficiency & \multicolumn{2}{c}{ 1.0 } \\
\hspace*{5mm} Photon identification efficiency & \multicolumn{2}{c}{ 2.0 } \\
\hspace*{5mm} Photon energy scale & \multicolumn{2}{c}{ Shape, 0.5 } \\
\hspace*{5mm} \ptmiss mismeasurement ($\Pg\Pg\Ph$ and VBF) & \NA & 50 \\
\hspace*{5mm} $\Delta\phi$ selection efficiency ($\Pg\Pg\Ph$ and VBF) & \NA & 1--4 \\
\end{tabular}
\end{table}

\subsection{The \texorpdfstring{\Htt}{h to tautau} channel}

The systematic uncertainties in the \Htt channel are related to the normalization of signal
and background processes and, in several instances, the shapes of the signal and background distributions.
As mentioned earlier, the simultaneous maximum-likelihood fit is performed in the SR and CRs,
where the shape and normalization uncertainties are represented by nuisance parameters in the likelihood.
Uncertainties affecting the distribution of \mttot (shape uncertainties) are represented by Gaussian pdfs, whereas log-normal pdfs are used for normalization, as stated above.
The largest overall uncertainty is statistical.
Table~\ref{tab:HttSystematics} summarizes the different sources of systematic uncertainty in this channel.

\begin{table}
   \topcaption{Systematic uncertainties affecting signal and background in the \Htt channel.}\label{tab:HttSystematics}
    \centering
      \cmsTable{
        \begin{tabular}{lcccc}
                &   & \multicolumn{3}{c}{Change in acceptance or shape}  \\ \cline{3-5}
            Source            & Affected processes  & $\Pe\tauh$ & $\Pgm\tauh$ & $\tauh\tauh$ \\
            \hline
            $\tauh$ identification (correlated)         & simulation & 4.5\% & 4.5\% & \NA \\
            $\tauh$ identification (uncorrelated)       & simulation & 2\% & 2\% & 9\% \\
            High \pt $\tauh$                            & simulation & \multicolumn{3}{c}{Shape, up to 8\%} \\
            $\Pe$ identification \& trigger          & simulation & 2\% & \NA & \NA \\
            $\Pgm$ identification \& trigger            & simulation & \NA & 2\% & \NA \\
            $\tauh$ trigger                             & simulation & \NA       & \NA     & Shape only \\
            $\Pe$ misidentified as $\tauh$          & $\PZ\to \Pe \Pe$    & 12\% & \NA & \NA \\
            $\Pgm$ misidentified as $\tauh$          & $\PZ\to \Pgm \Pgm$  & \NA & 25\% & \NA \\
            Jet misidentified as $\tauh$            & \zjet  & \multicolumn{3}{c}{Shape only} \\
            $\tauh$ energy scale (per decay mode)      & simulation & \multicolumn{3}{c}{1.2\% on energy scale} \\
            Jet energy scale and effect on $\ptmiss$    & simulation & \multicolumn{3}{c}{Shape, up to 10\%} \\
            \ptmiss energy scale                      & simulation & \multicolumn{3}{c}{Shape, up to 11\%} \\
            Integrated luminosity                  & simulation & \multicolumn{3}{c}{2.5\%} \\
            Norm.\ \wjet/QCD multijet         & \wjet/QCD multijet  & \multicolumn{3}{c}{up to 20\%} \\
            Norm.\ $\ttbar$                        & $\ttbar$       & \multicolumn{3}{c}{6\%} \\
            Norm.\ diboson                          & Diboson             & \multicolumn{3}{c}{5\%} \\
            Norm.\ single top                      & Single top          & \multicolumn{3}{c}{5\%} \\
            Norm.\ SM Higgs boson                      & SM Higgs boson      & \multicolumn{3}{c}{up to 5\%} \\
            \zjet LO-NLO reweighting              & \zjet   & \multicolumn{3}{c}{Shape, up to 26\%}\\
            \wjet NLO EW correction        & \wjet          & \multicolumn{3}{c}{Shape, up to 6\%}\\
            $\PW\PW$ NLO EW correction      & $\PW\PW$            & \multicolumn{3}{c}{Shape, up to 12\%}\\
            $\PZ\PZ$ NLO EW correction      & $\PZ\PZ$            & \multicolumn{3}{c}{Shape, up to 2\%}\\
            Top quark \pt reweighting                & $\ttbar$            & \multicolumn{3}{c}{Shape, up to 5\%}\\
            Theory: Higgs boson branching  & Signal + SM Higgs boson  & \multicolumn{3}{c}{1.7\%} \\
            \quad fraction &       & \\
            Theory: renorm. and fact. scale & Signal    & \multicolumn{3}{c}{4\%} \\
            Theory: PDF                          & Signal         & \multicolumn{3}{c}{ 2\%} \\
            Limited number of events & All processes      & \multicolumn{3}{c}{Shape only} \\
            \quad (bin-by-bin) &       & \\
        \end{tabular}
        }
\end{table}

An uncertainty of 2\% is assigned to simulated events containing an electron or muon candidate.
In simulated events with a $\tauh$ candidate, an additional uncertainty of 5\% per $\tauh$ is applied.
These uncertainties account for the observed differences in the performance of electron, muon, and $\tauh$ identification,
isolation, and trigger algorithms, between data and simulation.
The hadronic $\tauh$ efficiency is not fully correlated across all $\Pgt\Pgt$ final states because there are different discriminators used in each channel. The $\tauh\tauh$ channel has a 9\% $\tauh$ uncertainty due to a correlation with the $\tauh\tauh$ trigger systematic uncertainty.
An uncertainty of 12\% is assigned to simulated events containing an electron misidentified as a $\tauh$ candidate,
and 25\% for a muon misidentified as a $\tauh$ candidate~\cite{Sirunyan:2017khh}.
A 2 (4)\% uncertainty is assigned to the yield of multiboson and single top (\ttbar) processes to account for changes in overall
normalization arising from uncertainties in the \cPqb\ tagging performance.
Similarly, a 5\% \cPqb\ tagging uncertainty is assigned to \zjet and SM Higgs boson processes,
while all other processes, including signal, receive a 2\% uncertainty.
A systematic uncertainty of up to 20\% is applied to QCD multijet background to account for yield differences in the same-sign CR.
All of the background systematic uncertainties in the same-sign region are propagated to the total QCD multijet background uncertainty, which is taken to be 40\%.

The \wjet background has a \pt-dependent uncertainty, which approaches 10\%, from predicted NLO EW K-factors
where the full EW correction is treated as the systematic uncertainty~\cite{Sirunyan:2017hci,Kuhn:2005gv,Kallweit:2014xda,Kallweit:2015dum}.
Cross section uncertainties of the order of 5\% are applied to the $\ttbar$ (6\%), top quark (5\%), and diboson (5\%) processes~\cite{Sirunyan:2016cdg,Khachatryan:2016tgp,Sirunyan:2017zjc,Khachatryan:2016mnb}.
In simulated \zjet samples, a shape uncertainty of 10\% of the $\PZ$ boson \pt reweighting correction, to account for higher-order effects, is used.
The uncertainty in the \zjet background contribution is about 12\% in the SR.
The \ttbar contribution includes a shape systematic uncertainty equivalent to 5\% related to the top quark \pt spectrum, since there is evidence that the spectrum is softer in data than in simulation~\cite{Khachatryan:2016mnb}.

A 1.2\% uncertainty in the $\tau$ lepton energy scale~\cite{Sirunyan:2017khh} is propagated through to the final signal extraction variables.
The $\tau$ lepton energy scale depends on the \tauh decay mode and is correlated across all channels.
A shape uncertainty is used for the uncertainty in the double \tauh trigger.
A shift of 3\% of the \pt of the trigger-level \tauh candidate leads to a 12\% normalization difference at 40\GeV, and a 2\% difference at 60\GeV.
For $\pt>60\GeV$, a constant 2\% systematic uncertainty is applied.

To account for potentially different rates of jets misidentified as \tauh candidates between data and simulation,
an uncertainty, applied as a function of the \pt of the \tauh candidate,
is used for background events where the reconstructed \tauh candidate is matched to a jet at generator-level.
The uncertainty increases to about 20\% near a \tauh candidate $\pt=200\GeV$, and acts to change the shape of the \mttot distribution.
An asymmetric uncertainty related to the identification of \tauh with a high \pt is applied to signal and background simulation. The high-\pt \tauh efficiency measurement uses selected highly virtual \PW\xspace bosons and has limited statistical precision in comparison to the lower \pt $\PZ\to\tau\tau$ and \ttbar \tauh efficiency studies. Therefore the asymmetric uncertainty is used in combination with a constant scale factor.
It is proportional to \pt and has a value of $+5\%$ and $-35\%$ per 1\TeV.
For the application of all of the aforementioned $\Pgt$ lepton uncertainties, simulated backgrounds are separated depending on whether the reconstructed \tauh candidates are matched to generated $\Pgt$ leptons.

In all simulated samples, uncertainties in the \ptmiss calculation related to unclustered energy deposits are taken into account.
Uncertainties in the jet energy scale are included on an event-by-event basis and propagated to the \ptmiss calculation.
Lastly, an uncertainty in the statistical precision of each process in each bin of the distribution is also included.

\section{Results \label{sec:Results}}

The results of the analysis are derived from the maximum-likelihood fits presented in Sections~\ref{sec:AnalysisHgg} and \ref{sec:AnalysisHtt} for the \Hgg and \Htt channels, respectively.

\subsection{Observed yields \label{sec:Yields}}

For the \Hgg channel, from the signal plus background fit to \mgg, the number of expected
events from background processes are determined.
The background yields and the observed number of events within 3\GeV of the SM Higgs boson mass are listed in Table~\ref{tab:HggYields} for both the low- and high-\ptmiss
categories.
The excess at low-\ptmiss has negligible effect on the results when combined with the high-\ptmiss category for the benchmark signals considered in this paper.

\begin{table}[ht]
\centering
\topcaption{\label{tab:HggYields}
Expected background yields and observed numbers of events for the \Hgg channel in the \mgg range of 122--128\GeV
are shown for the low- and high-\ptmiss categories.
The nonresonant background includes QCD multijet, $\gamma\gamma$, \GAMJET, and EW backgrounds and is estimated from the analytic function fit to data.
The SM Higgs boson background is presented separately for the irreducible V\Ph production and for the other production modes.
For the resonant background contributions, both the statistical and the systematic uncertainties are listed.
As detailed in Section~\ref{sec:HggExtraction}, the systematic uncertainty associated with the nonresonant background is negligible.}
\begin{tabular}{lcc}
    Expected background & Low-\ptmiss category & High-\ptmiss category \\
    \hline
    SM \Hgg (V\Ph)                              & $2.9 \pm 0.1\stat \pm 0.2\syst$               & $1.26 \pm 0.05\stat\pm 0.09\syst$ \\
    SM \Hgg ($\Pg\Pg\Ph$, $\ttbar$\Ph, VBF)     & $5.3 \pm 0.3\stat \pm 1.2\syst$               & $0.11 \pm 0.01\stat\pm 0.01\syst$ \\
    Nonresonant background                      & $125.1 \pm 11.2\stat$                         & $4.5 \pm 2.1\stat$                \\[\cmsTabSkip]
    Total background                            & $133 \pm 11\stat \pm 1\syst$                  & $5.9 \pm 2.1\stat \pm 0.1\syst$ \\[\cmsTabSkip]
    Observed events                                     & 159                                           & 6 \\
\end{tabular}
\end{table}

In the \Htt channel, the final simultaneous fit to the \mttot distributions for the SR, and \wjet and QCD multijet CRs
is performed in each of the three considered $\Pgt$ decay channels (\Pe\tauh, \Pgm\tauh, and $\tauh\tauh$).
The extracted post-fit yields for the expected number of background events and the number of events observed in data
are shown in Table~\ref{tab:HttYields}.
The number of events observed is in good agreement with the number of events predicted by the SM backgrounds.

\begin{table}[ht]
    \centering
    \topcaption{
        Estimated background yields and observed numbers of events for $\mttot >260\GeV$, in the SR of the \Htt channel.
        The uncertainties in the total expected yields include the statistical and systematic contributions.
    }
    \label{tab:HttYields}
    \begin{tabular}{lccc}

        Expected background             & $\Pe\tauh$       & $\Pgm\tauh$        & $\tauh\tauh$  \\
        \hline
        \wjet /QCD multijet    & 13.1 $\pm$ 2.2      & 32.5 $\pm$ 6.2   & 3.8  $\pm$ 2.6 \\
        $\ttbar$                & 13.7 $\pm$ 1.6   & 24.8 $\pm$ 2.0   & 4.2  $\pm$ 1.3 \\
        SM Higgs boson          & 0.48  $\pm$ 0.08  & 0.72  $\pm$ 0.06  & 1.21  $\pm$ 0.08 \\
        Diboson                 & 12.3 $\pm$ 1.0    & 21.5 $\pm$ 1.5     & 7.3  $\pm$ 0.6 \\
        $\PZ\to\Pgt\Pgt$    & 0.00  $\pm$ 0.01    & 0.0  $\pm$ 0.5  & 3.6  $\pm$ 1.2 \\
        $\PZ\to\ell\ell$    & 0.9  $\pm$ 1.9     & 2.0  $\pm$ 1.3   &  \NA               \\
        $\PZ\to\nu\nu$      & \NA                 & \NA                 & 0.4  $\pm$ 0.3 \\[\cmsTabSkip]
        Total background          & 40.5 $\pm$ 3.3    & 81.8 $\pm$ 6.3   & 20.5 $\pm$ 3.0 \\[\cmsTabSkip]
        Observed events           & 38    &  81   & 26  \\
    \end{tabular}
\end{table}

Aside from the small excess in the low-\ptmiss category of the \Hgg channel,
the observed numbers of events are consistent with SM expectations.
All of the results presented here are interpreted in terms of the two benchmark models of DM production mentioned earlier.
Expected signal yields and the product of the predicted signal acceptances and their efficiencies (\Ae)
are summarized in Table~\ref{tab:SigYieldEff} for selected mass points, in both \Hgg and \Htt channels.

\begin{table}[ht]
\centering
\topcaption{
    The expected signal yields and the product of acceptance and efficiency (\Ae) for the two benchmark models.
    The \PZpr-2HDM signal is shown for the parameters \mA\xspace = 300\GeV and \mZ\xspace = 1000\GeV, and the baryonic \PZpr signal, for the parameters \mDM\xspace = 1\GeV and \mZ\xspace = 100\GeV.
\label{tab:SigYieldEff}}
\begin{tabular}{lcccccc}
    & \multicolumn{2}{c}{\Hgg channel}             & & \multicolumn{3}{c}{\Htt channel}  \\
    Signal & Low-\ptmiss & High-\ptmiss  & & \Pe\tauh & \Pgm\tauh & \tauh\tauh \\
    \hline
    \PZpr-2HDM & & & & & & \\
    \xspace\xspace\xspace\xspace Expected yield	& 0.1 $\pm$ 0.4 	& 4.5 $\pm$ 0.6	 	& & 6.5 $\pm$ 0.3     & 11.1 $\pm$ 0.5 	& 14.3 $\pm$ 1.2 \\
    \xspace\xspace\xspace\xspace \Ae [\%]	& 0.1 			& 42.6  		& & 2.2			& 3.6			& 4.4	 \\
    Baryonic \PZpr & & & & & & \\
    \xspace\xspace\xspace\xspace Expected yield	& 14.7 $\pm$ 6.7	& 13.8 $\pm$ 6.4 	& & 8.6 $\pm$ 0.3     & 16.8 $\pm$ 0.5 	& 20.9 $\pm$ 0.8\\
    \xspace\xspace\xspace\xspace \Ae [\%]	& 6.4 			& 6.0 			& & 0.1			& 0.3			& 0.3	 \\
\end{tabular}
\end{table}

A discussion of the results for the \PZpr-2HDM interpretation is presented in Section~\ref{sec:2HDMResults}.
The results in the context of the baryonic \PZpr interpretation are given in Section~\ref{sec:BaryonicResults}.
The baryonic \PZpr results are also reinterpreted for comparison with direct detection experiments in Section~\ref{sec:SpinIndResults}
by looking at simplified DM models proposed by the ATLAS-CMS Dark Matter Forum~\cite{darkMatterForum}.

\subsection{Interpretation in the \texorpdfstring{\PZpr}{Z'}-2HDM model \label{sec:2HDMResults}}

For the event selection given in Sections~\ref{sec:AnalysisHgg} and \ref{sec:AnalysisHtt},
the results interpreted in terms of the \PZpr-2HDM associated production of DM and a Higgs boson are presented here.
The expected and observed yields are used to calculate an upper limit on the production cross section of DM+\Ph production via the \PZpr-2HDM mechanism.
Upper limits are computed~\cite{combineProcedure} at 95\% confidence level (\CL) using a profile likelihood ratio
and the modified frequentist criterion~\cite{Junk,Read2}
with an asymptotic approximation~\cite{CLS3}.
The upper limits are obtained for each Higgs boson decay channel separately and for the statistical combination of the two.
The two decay channels are combined using the Higgs boson branching fractions predicted by the SM~\cite{deFlorian:2016spz}.
In the combination of the two analyses, the theoretical uncertainties in the Higgs boson cross section and
the systematic uncertainty in the integrated luminosity are assumed to be fully correlated between the two decay channels.

\begin{figure}
  \centering
  \includegraphics[width=0.7\textwidth]{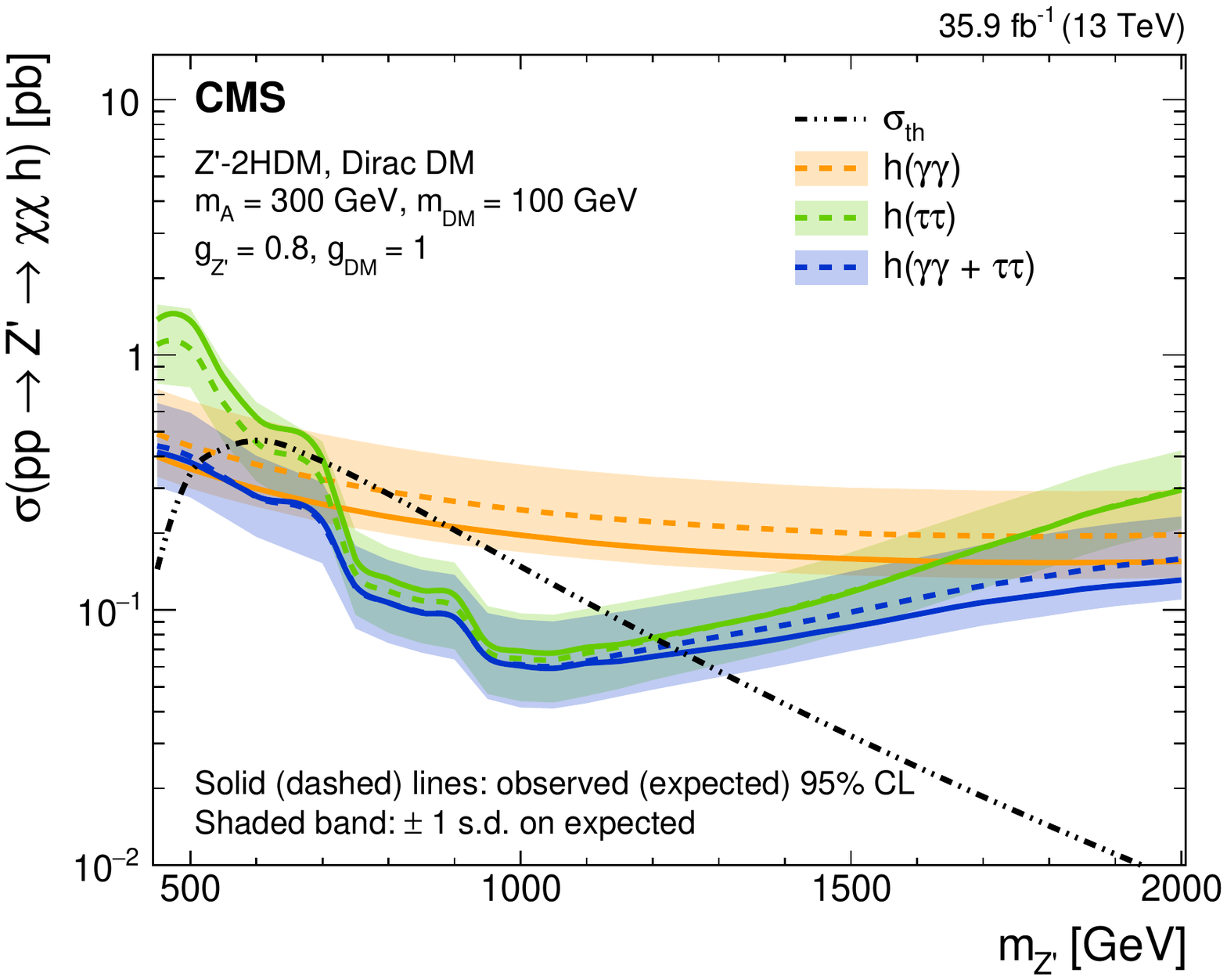}
  \caption{\label{fig:2HDM_UL_XSEC}
Expected and observed 95\% \CL upper limits on the \PZpr-2HDM cross section for dark matter associated production with a Higgs boson ($\PZpr\to\PGc\PGc\Ph$) are shown.
Limits are given for the \Hgg channel, \Htt channel, and their combined exclusion.
  }
\end{figure}

Figure~\ref{fig:2HDM_UL_XSEC} shows the 95\% \CL expected and observed upper limits
on the DM production cross section (\SigCL) as a function of \PZpr mass.
Both the \Hgg and \Htt channels, as well as the combination of the two, are shown for $\mA = 300\GeV$.
These upper limits, although obtained with a DM mass of 100\GeV, can be considered valid for any DM mass below 100\GeV
since the branching fraction for decays of \PSA to DM particles decreases as the dark matter mass increases.
The theoretical cross section (\SigTH) is calculated with $\mDM\xspace = 100\GeV$, $\gZp = 0.8$,
and $\gDM = \tan{\beta} = 1$, as mentioned in Section~\ref{sec:Introduction}.

To produce exclusion limits in the two-dimensional plane of \PZpr mass and \PSA mass, an interpolation is performed.
Fully simulated signal samples (mentioned in Section~\ref{sec:Samples}) were generated in a coarse grid of \mA\xspace and \mZ.
For the \Hgg channel, the \mgg shape does not depend on the mass of these particles, only the expected yield is affected by these masses.
Therefore, the product \Ae of the fully simulated samples is parametrized and used to extract the expected number of events for intermediate mass points.
In the \Htt channel, this is not sufficient because the \mttot shape does depend on the particle masses.
A reweighting technique is used to extract the yields for the intermediate mass points.
Simulation samples were produced at generator-level for \mZ\xspace between 450 and 2000\GeV in steps of 50\GeV and for \mA\xspace between 300 and 700\GeV in steps of 25\GeV.
These are compared with the full-simulation samples at generator-level.
The bin-by-bin ratio of the SM-like Higgs boson \pt between the two samples is used to weight the full-simulation samples.
This method was validated by applying the same procedure at the generator-level among the samples for which full-simulation is available.

\begin{figure}
  \centering
  \includegraphics[width=0.49\textwidth]{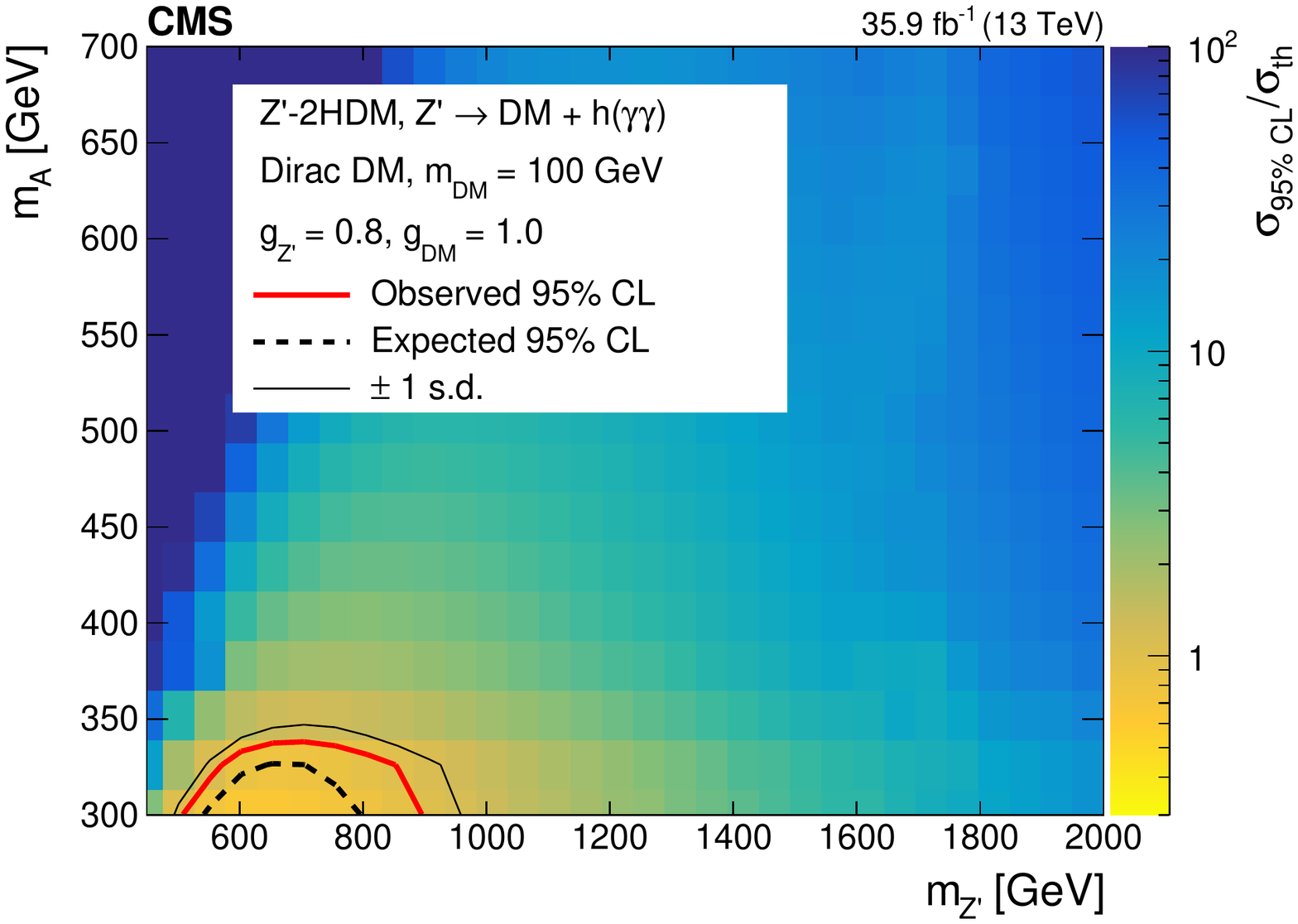}
  \includegraphics[width=0.49\textwidth]{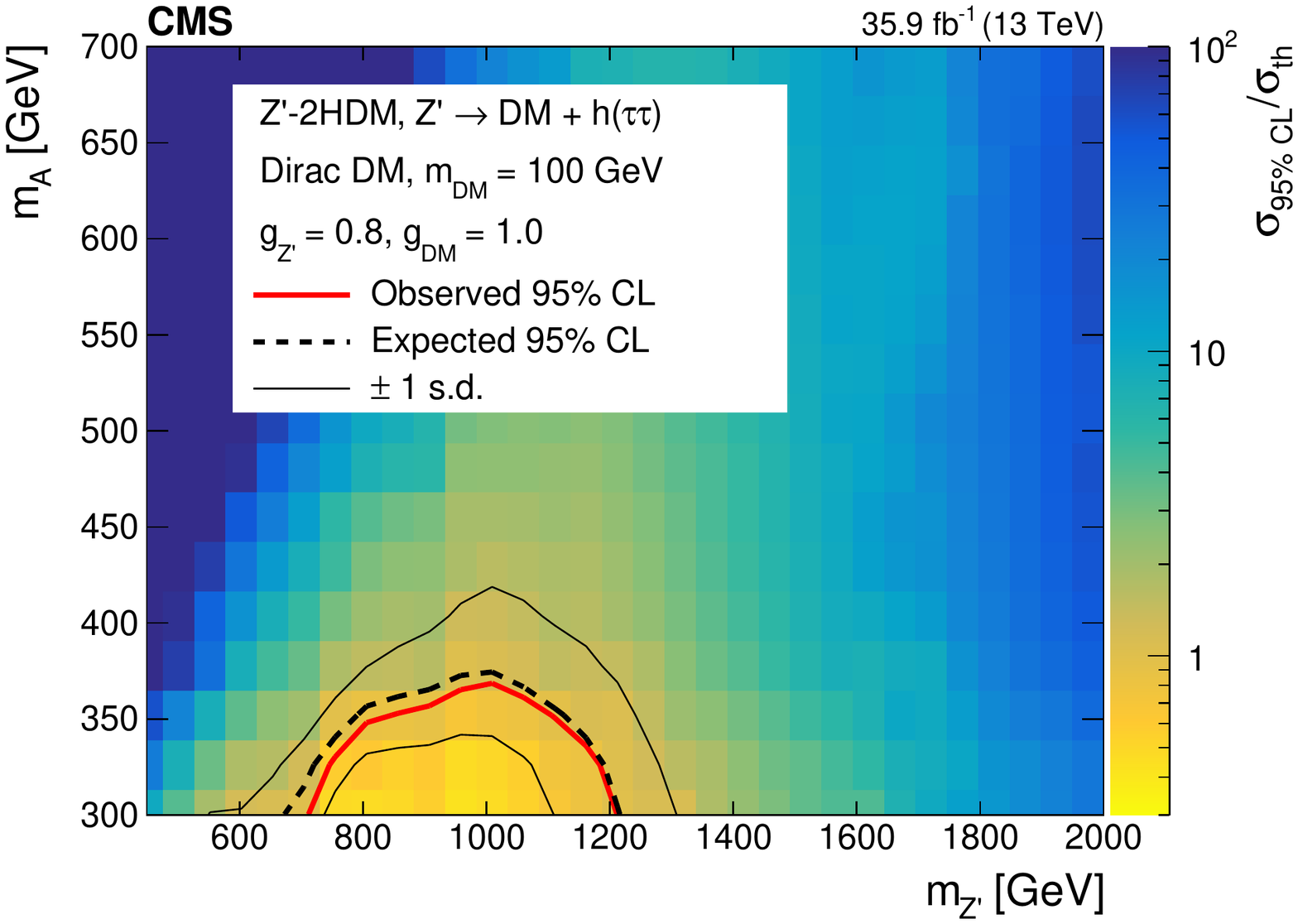}
  \includegraphics[width=0.49\textwidth]{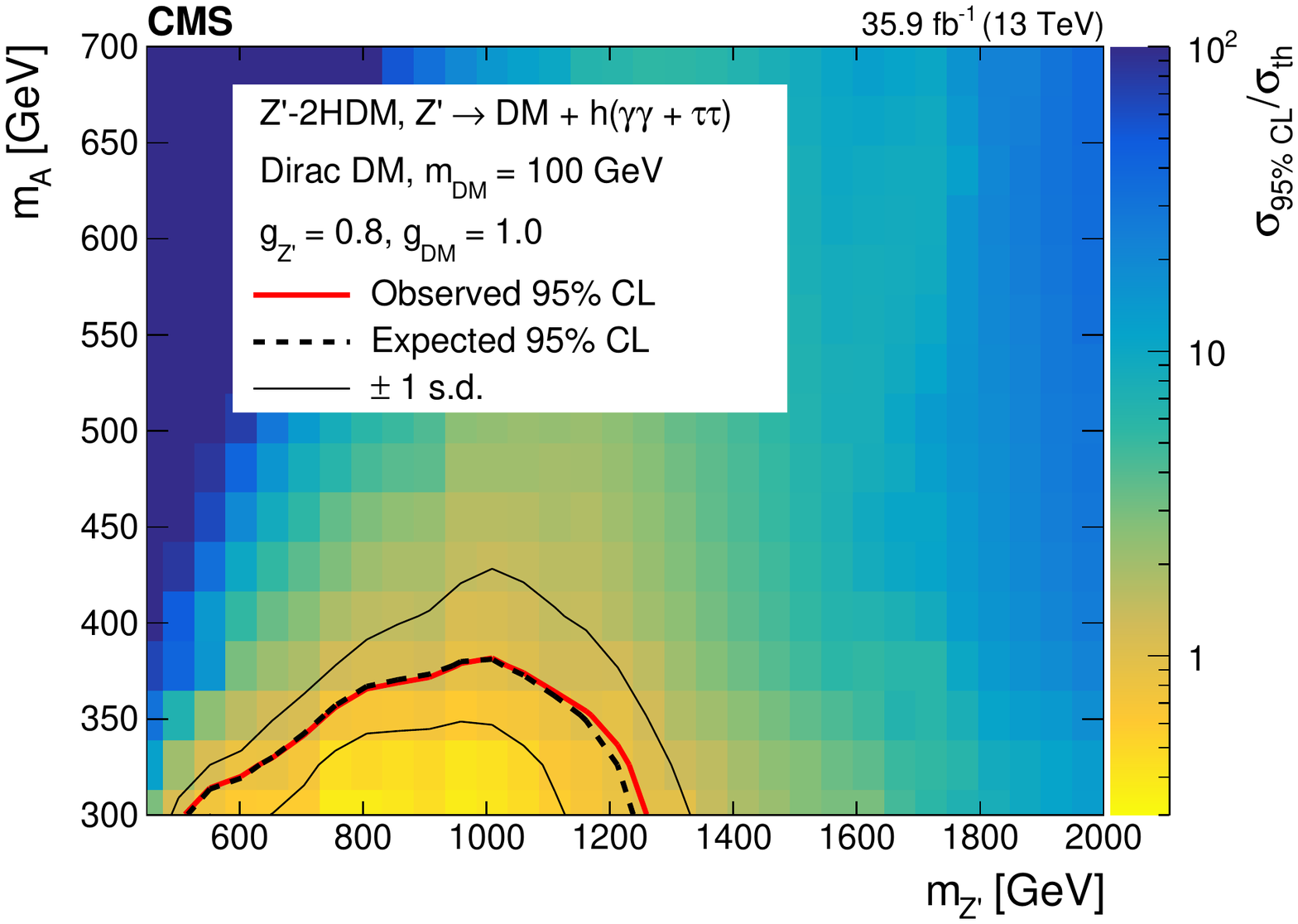}
  \caption{\label{fig:2HDM_UL_MU}
  Observed 95\% \CL upper limits on the \PZpr-2HDM signal strength for the \Hgg (left), \Htt (right), and combination of the two channels (lower center).
  The observed (expected) two-dimensional exclusion curves are shown with thick red (dashed black) lines.
  The plus and minus one standard deviation expected exclusion curves are also shown as thin black lines.
  The region below the lines is excluded.
  }
\end{figure}

The interpolation between mass points is improved using kernel algorithms to display smooth, continuous exclusion contours.
The resulting two-dimensional exclusion for the \PZpr-2HDM signal is shown in Fig.~\ref{fig:2HDM_UL_MU}.
The 95\% \CL expected and observed upper limits on signal strength (\SigCL/\SigTH) are shown.
Regions of the parameter space with $\SigCL/\SigTH < 1$ are excluded at 95\% \CL under the nominal \SigTH hypothesis.
For $\mA = 300\GeV$, the \Hgg channel alone excludes at 95\% \CL the \PZpr masses from 550\GeV to 860\GeV,
while the \Htt channel excludes the \mZ\xspace masses from 750\GeV to 1200\GeV.
The combination of these two decay channels excludes the \PZpr masses from 550\GeV to 1265\GeV for $\mA = 300\GeV$.
The \PZpr mass range considered is extended from previous CMS searches to include $450 \leq \mZ < 600 \GeV$.

\subsection{Baryonic \texorpdfstring{\PZpr}{Z'} model interpretation \label{sec:BaryonicResults}}

\begin{figure}
  \centering
  \includegraphics[width=0.7\textwidth]{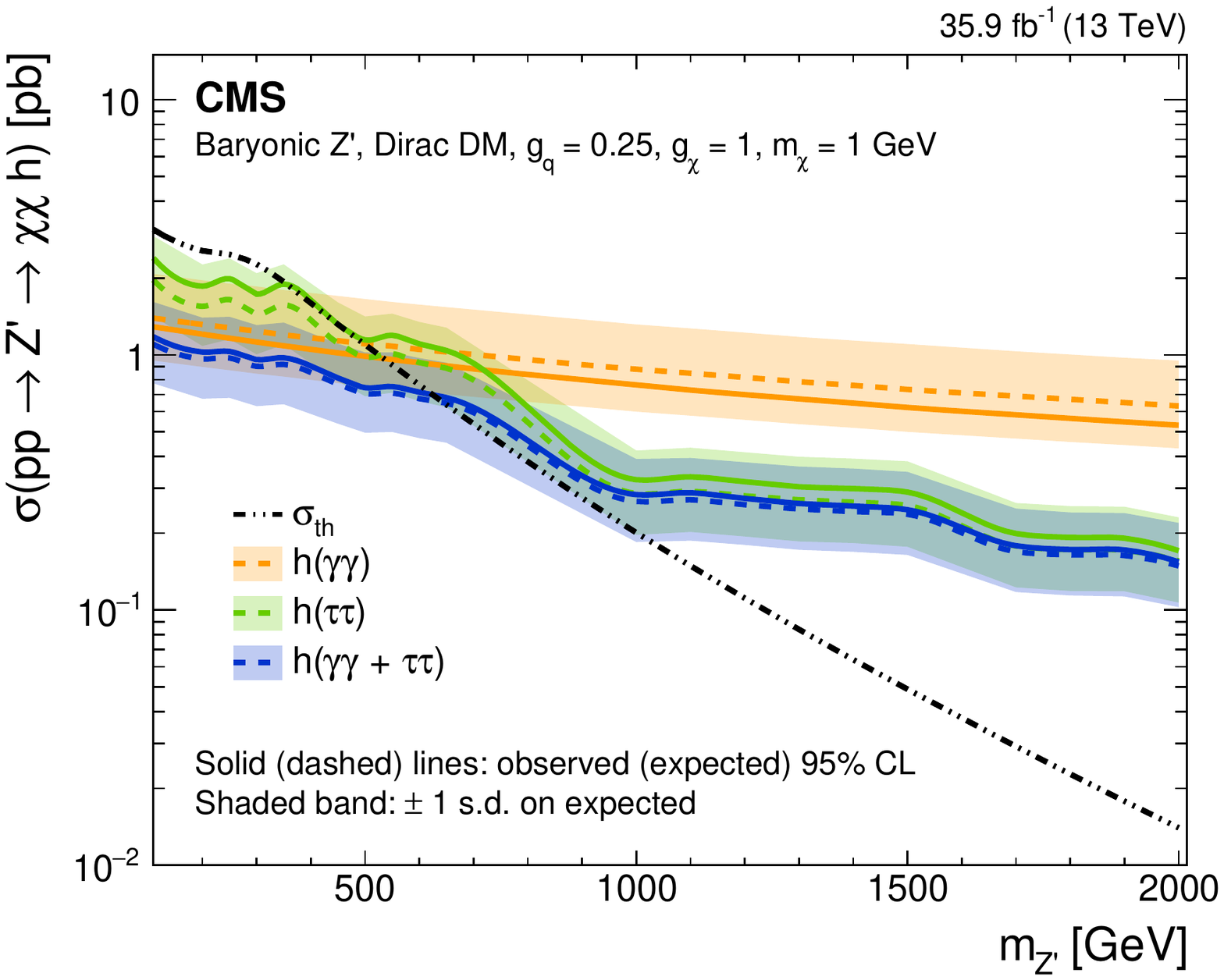}
  \caption{\label{fig:BARY_UL_XSEC}
Expected and observed 95\% \CL upper limits on the baryonic \PZpr cross section for dark matter associated production with a Higgs boson ($\PZpr\to\PGc\PGc\Ph$) are shown.
Limits are given for the \Hgg channel, \Htt channel, and their combined exclusion.
  }
\end{figure}

\begin{figure}
  \centering
  \includegraphics[width=0.49\textwidth]{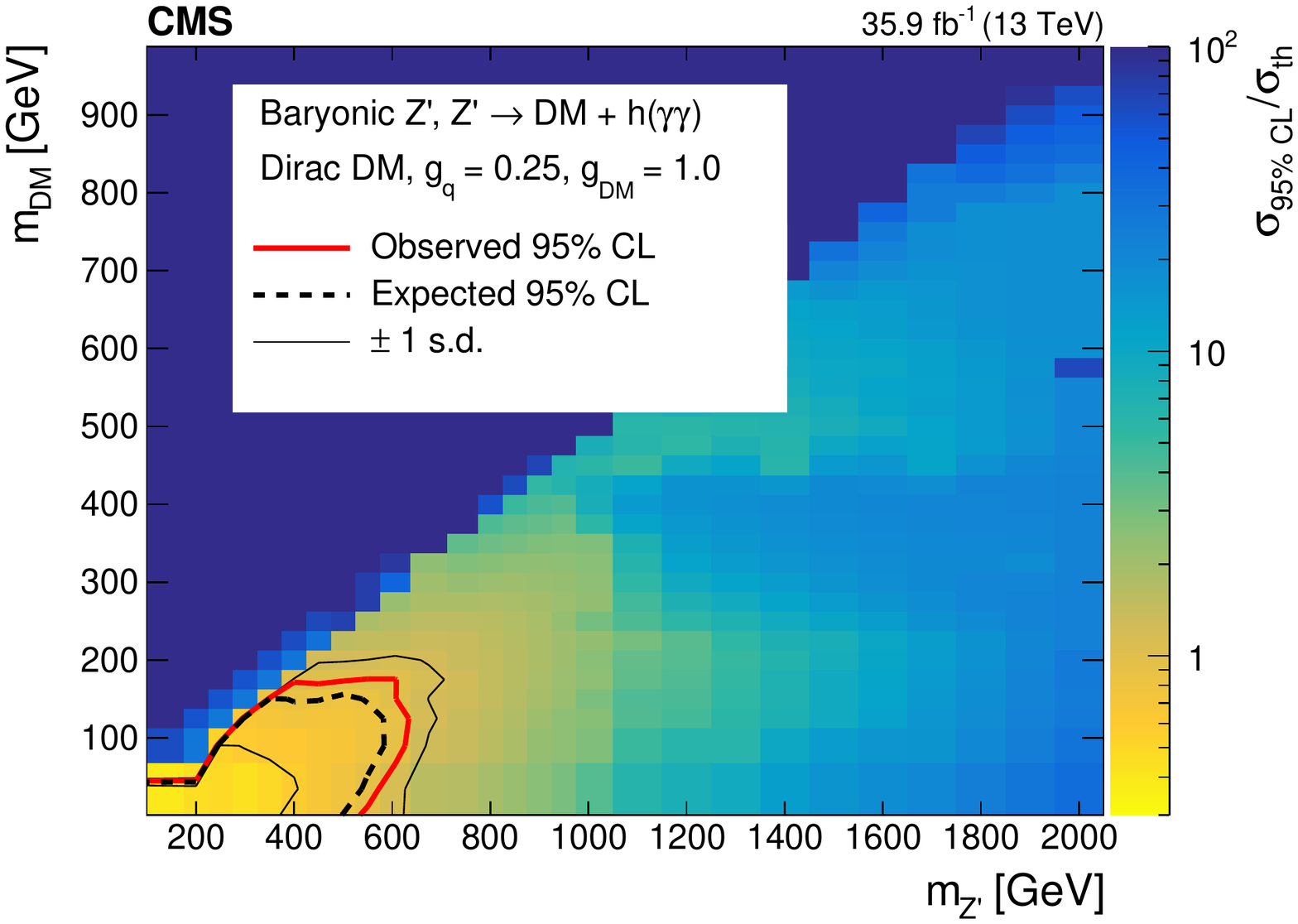}
  \includegraphics[width=0.49\textwidth]{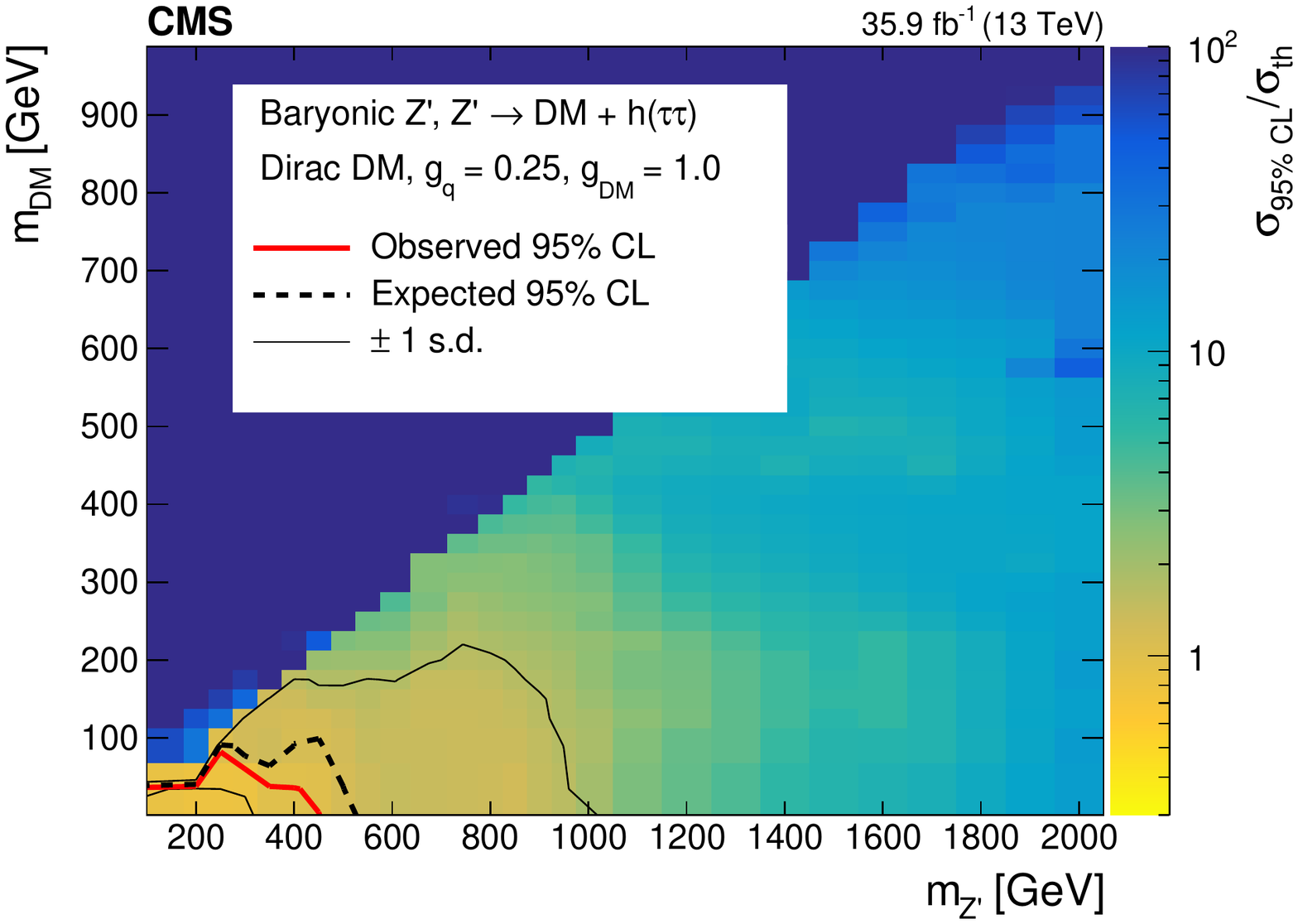}
  \includegraphics[width=0.49\textwidth]{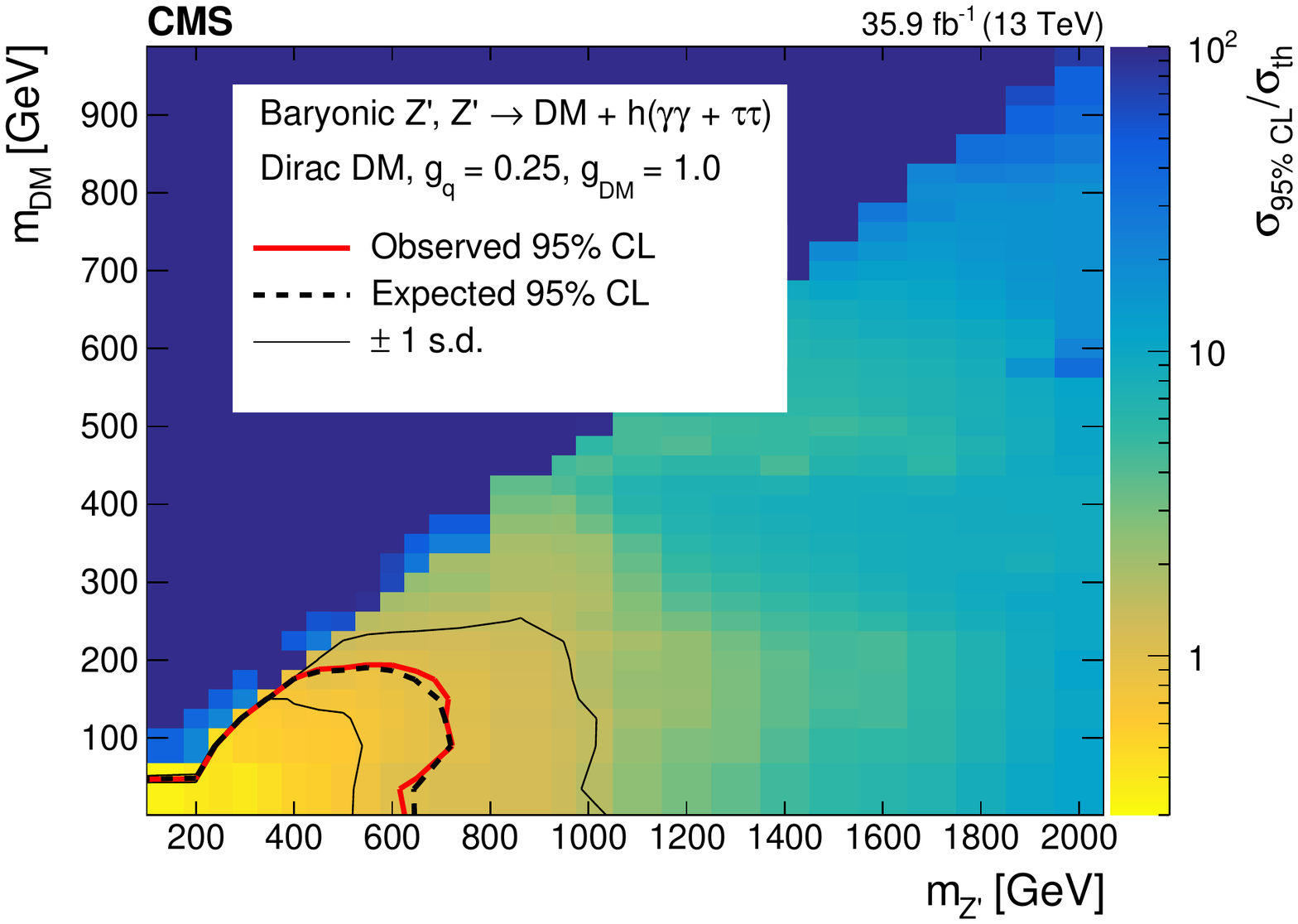}
  \caption{\label{fig:BARY_UL_MU}
  Observed 95\% \CL upper limits on the baryonic \PZpr signal strength for the \Hgg (left), \Htt (right), and combination of the two channels (lower center).
  The observed (expected) two-dimensional exclusion curves are shown with thick red (dashed black) lines.
  The plus and minus one standard deviation expected exclusion curves are also shown as thin black lines.
  The region below the lines is excluded.
  }
\end{figure}

Here the results presented in Section~\ref{sec:Yields} are interpreted in the context of the baryonic \PZpr model.
This paper presents the first baryonic \PZpr model interpretation of $\Ph + \ptmiss$ searches with the CMS detector.
The 95\% \CL upper limits on DM+\Ph cross section are calculated for the baryonic \PZpr production mechanism.
The upper limits for each decay channel and the combination of the two channels are shown in Fig.~\ref{fig:BARY_UL_XSEC}.
The \SigTH is calculated assuming the choice of parameters detailed in Section~\ref{sec:Introduction}.
Results in the two-dimensional plane of \mDM\xspace and \mZ\xspace are produced using an interpolated grid
produced in the same way as described in Section~\ref{sec:2HDMResults}.
The two-dimensional exclusion for this model is shown in Fig.~\ref{fig:BARY_UL_MU}, where the 95\% \CL upper limits on the signal strength are shown
for each decay channel and for the combination of the \Hgg and \Htt channels.
For $\mDM = 1\GeV$, the \Hgg channel excludes \mZ\xspace masses up to 574\GeV.
The \Htt channel similarly excludes \mZ\xspace masses up to 450\GeV.
The combination of the two decay channels excludes \mZ\xspace up to 615\GeV for $\mDM = 1\GeV$.

\subsection{Simplified DM model interpretation \label{sec:SpinIndResults}}

Limits from the baryonic \PZpr model are reinterpreted to infer limits on the $s$-channel simplified DM models
that were proposed by the ATLAS-CMS Dark Matter Forum~\cite{darkMatterForum} for comparison with direct detection experiments.
In the model considered in this analysis, Dirac DM particles couple to a vector mediator, which in turn couples to the SM quarks.
A point in the parameter space of this model is determined by four variables: the DM particle mass \mDM, the mediator mass \mMed,
the mediator-DM coupling \gDM, and the universal mediator-quark coupling \gq.
The couplings for this analysis are fixed to $\gDM = 1.0$ and $\gq = 0.25$, following the recommendation of Ref.~\cite{presentDM}.

\begin{figure}
\centering
\includegraphics[width=0.9\textwidth]{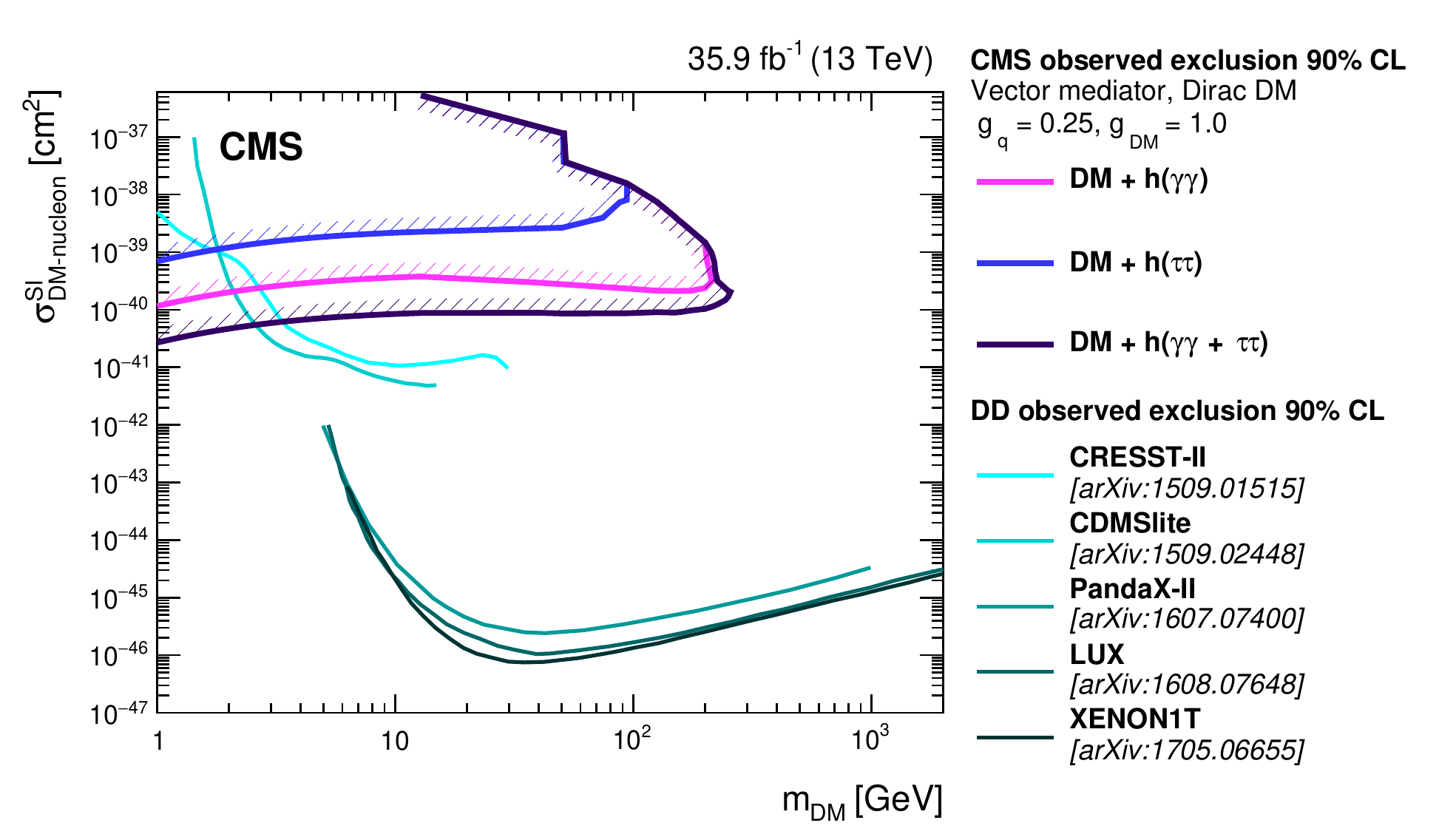}
\caption{  \label{fig:Xsec_mDM}
The 90\% \CL exclusion limits on the DM-nucleon SI scattering cross section as a function of \mDM.
Results obtained in this analysis are compared with those from a selection of direct detection (DD) experiments.
The latter exclude the regions above the curves.
Limits from CDMSLite~\cite{CDMSLite}, LUX~\cite{LUX}, XENON-1T~\cite{XENON1T}, PandaX-II~\cite{PandaxII}, and CRESST-II~\cite{CresstII} are shown.
}
\end{figure}

The results are interpreted in the spin-independent (SI) cross section \SigSI for DM scattering off a nucleus.
The value of \SigSI for a given point in the $s$-channel simplified DM model is determined by the equation~\cite{presentDM}:
\begin{equation}
\SigSI = \frac{f^2(\gq)\gDM^2\mu^2_{\mathrm{nDM}}}{\pi \mMed^4},
\end{equation}
where $\mu_{\mathrm{nDM}}$ is the reduced mass of the DM-nucleon system and $f(\gq)$ is the mediator-nucleon coupling, which is dependent on \gq.
The resulting \SigSI limits as a function of DM mass are shown in Fig.~\ref{fig:Xsec_mDM}.
In the same plot, exclusions from several direct detection experiments are shown.
For the baryonic \PZpr model, the limits are more stringent than direct detection
experiments for $\mDM < 2.5\GeV$.

\section{Summary \label{sec:Conclusion}}

A search for dark matter particles produced in association with a Higgs boson has been performed.
The study focuses on the case where the 125\GeV Higgs boson decays to either two photons or two $\tau$ leptons.
This analysis is based on proton-proton collision data collected with the CMS detector during 2016 at $\sqrt{s}$ = 13\TeV,
corresponding to an integrated luminosity of 35.9\fbinv.
The results of the search are interpreted in terms of a \PZpr-two-Higgs-doublet model (\PZpr-2HDM) and a baryonic \PZpr simplified model of dark matter production.

A statistical combination of the two channels was performed and these results were used to produce upper limits on dark matter production.
Limits on the signal production cross section are calculated for both simplified models.
For the \PZpr-2HDM signal, with an intermediate pseudoscalar of mass $\mA = 300\GeV$ and $\mDM = 100\GeV$,
the \PZpr masses from 550\GeV to 1265\GeV are excluded at 95\% confidence level.
For the baryonic \PZpr model, with $\mDM = 1\GeV$, \PZpr masses up to 615\GeV are excluded.
This is the first search for dark matter produced in association with a Higgs boson decaying to two $\tau$ leptons
and the first to combine results from the $\gamma\gamma$ and $\Pgt^+\Pgt^-$ decay channels.
The \PZpr-2HDM interpretation extended the \PZpr mass range compared with previous CMS searches.
The interpretation of the results include the baryonic \PZpr model interpretation for CMS
and an extrapolation to limits on the spin-independent cross section for the dark matter-nucleon interaction.

\begin{acknowledgments}
We congratulate our colleagues in the CERN accelerator departments for the excellent performance of the LHC and thank the technical and administrative staffs at CERN and at other CMS institutes for their contributions to the success of the CMS effort. In addition, we gratefully acknowledge the computing centers and personnel of the Worldwide LHC Computing Grid for delivering so effectively the computing infrastructure essential to our analyses. Finally, we acknowledge the enduring support for the construction and operation of the LHC and the CMS detector provided by the following funding agencies: BMBWF and FWF (Austria); FNRS and FWO (Belgium); CNPq, CAPES, FAPERJ, FAPERGS, and FAPESP (Brazil); MES (Bulgaria); CERN; CAS, MoST, and NSFC (China); COLCIENCIAS (Colombia); MSES and CSF (Croatia); RPF (Cyprus); SENESCYT (Ecuador); MoER, ERC IUT, and ERDF (Estonia); Academy of Finland, MEC, and HIP (Finland); CEA and CNRS/IN2P3 (France); BMBF, DFG, and HGF (Germany); GSRT (Greece); NKFIA (Hungary); DAE and DST (India); IPM (Iran); SFI (Ireland); INFN (Italy); MSIP and NRF (Republic of Korea); MES (Latvia); LAS (Lithuania); MOE and UM (Malaysia); BUAP, CINVESTAV, CONACYT, LNS, SEP, and UASLP-FAI (Mexico); MOS (Montenegro); MBIE (New Zealand); PAEC (Pakistan); MSHE and NSC (Poland); FCT (Portugal); JINR (Dubna); MON, RosAtom, RAS, RFBR, and NRC KI (Russia); MESTD (Serbia); SEIDI, CPAN, PCTI, and FEDER (Spain); MOSTR (Sri Lanka); Swiss Funding Agencies (Switzerland); MST (Taipei); ThEPCenter, IPST, STAR, and NSTDA (Thailand); TUBITAK and TAEK (Turkey); NASU and SFFR (Ukraine); STFC (United Kingdom); DOE and NSF (USA).

\hyphenation{Rachada-pisek} Individuals have received support from the Marie-Curie program and the European Research Council and Horizon 2020 Grant, contract No. 675440 (European Union); the Leventis Foundation; the A. P. Sloan Foundation; the Alexander von Humboldt Foundation; the Belgian Federal Science Policy Office; the Fonds pour la Formation \`a la Recherche dans l'Industrie et dans l'Agriculture (FRIA-Belgium); the Agentschap voor Innovatie door Wetenschap en Technologie (IWT-Belgium); the F.R.S.-FNRS and FWO (Belgium) under the ``Excellence of Science - EOS" - be.h project n. 30820817; the Ministry of Education, Youth and Sports (MEYS) of the Czech Republic; the Lend\"ulet (``Momentum") Program and the J\'anos Bolyai Research Scholarship of the Hungarian Academy of Sciences, the New National Excellence Program \'UNKP, the NKFIA research grants 123842, 123959, 124845, 124850 and 125105 (Hungary); the Council of Science and Industrial Research, India; the HOMING PLUS program of the Foundation for Polish Science, cofinanced from European Union, Regional Development Fund, the Mobility Plus program of the Ministry of Science and Higher Education, the National Science Center (Poland), contracts Harmonia 2014/14/M/ST2/00428, Opus 2014/13/B/ST2/02543, 2014/15/B/ST2/03998, and 2015/19/B/ST2/02861, Sonata-bis 2012/07/E/ST2/01406; the National Priorities Research Program by Qatar National Research Fund; the Programa Estatal de Fomento de la Investigaci{\'o}n Cient{\'i}fica y T{\'e}cnica de Excelencia Mar\'{\i}a de Maeztu, grant MDM-2015-0509 and the Programa Severo Ochoa del Principado de Asturias; the Thalis and Aristeia programs cofinanced by EU-ESF and the Greek NSRF; the Rachadapisek Sompot Fund for Postdoctoral Fellowship, Chulalongkorn University and the Chulalongkorn Academic into Its 2nd Century Project Advancement Project (Thailand); the Welch Foundation, contract C-1845; and the Weston Havens Foundation (USA).

\end{acknowledgments}

\bibliography{auto_generated}

\appendix

\cleardoublepage \appendix\section{The CMS Collaboration \label{app:collab}}\begin{sloppypar}\hyphenpenalty=5000\widowpenalty=500\clubpenalty=5000\vskip\cmsinstskip
\textbf{Yerevan Physics Institute, Yerevan, Armenia}\\*[0pt]
A.M.~Sirunyan, A.~Tumasyan
\vskip\cmsinstskip
\textbf{Institut f\"{u}r Hochenergiephysik, Wien, Austria}\\*[0pt]
W.~Adam, F.~Ambrogi, E.~Asilar, T.~Bergauer, J.~Brandstetter, M.~Dragicevic, J.~Er\"{o}, A.~Escalante~Del~Valle, M.~Flechl, R.~Fr\"{u}hwirth\cmsAuthorMark{1}, V.M.~Ghete, J.~Hrubec, M.~Jeitler\cmsAuthorMark{1}, N.~Krammer, I.~Kr\"{a}tschmer, D.~Liko, T.~Madlener, I.~Mikulec, N.~Rad, H.~Rohringer, J.~Schieck\cmsAuthorMark{1}, R.~Sch\"{o}fbeck, M.~Spanring, D.~Spitzbart, A.~Taurok, W.~Waltenberger, J.~Wittmann, C.-E.~Wulz\cmsAuthorMark{1}, M.~Zarucki
% ...
\vskip\cmsinstskip
% ...
1:  Also at Vienna University of Technology, Vienna, Austria\\
% ...
\end{sloppypar}
\end{document}